\documentclass[sigconf,nonacm]{acmart}
\usepackage[T1]{fontenc}

\usepackage{array}
\usepackage{makecell}
\usepackage{amsmath}
\usepackage{graphicx}
\usepackage{xcolor}
\usepackage{enumitem}

\newcommand{\descr}[1]{ \noindent \textbf{#1}}

\usepackage{listings}
\usepackage{xcolor}

\begin{document}

\title{SpaLLM-Guard: Pairing SMS Spam Detection Using Open-source and Commercial LLMs}

\author{Muhammad Salman}
\affiliation{%
  \institution{Macquarie University}
  \city{Sydney}
  \country{Australia}
}
\email{muhammad.salman@mq.edu.au}

\author{Muhammad Ikram}
\affiliation{%
  \institution{Macquarie University}
  \city{Sydney}
  \country{Australia}
}
\email{muhammad.ikram@mq.edu.au}

\author{Nardine Basta}
\affiliation{%
  \institution{Macquarie University}
  \city{Sydney}
  \country{Australia}
}
\email{nardine.basta@mq.edu.au}

\author{Mohamed Ali Kaafar}
\affiliation{%
  \institution{Macquarie University}
  \city{Sydney}
  \country{Australia}
}
\email{dali.kaafar@mq.edu.au}

\renewcommand{\shortauthors}{Salman et al.}

\begin{abstract}

The increasing threat of SMS spam, driven by evolving adversarial techniques and concept drift, calls for more robust and adaptive detection methods. In this paper, we evaluate the potential of large language models (LLMs), both open-source and commercial, for SMS spam detection, comparing their performance across zero-shot, few-shot, fine-tuning, and chain-of-thought prompting approaches. Using a comprehensive dataset of SMS messages, we assess the spam detection capabilities of prominent LLMs such as GPT-4, DeepSeek, LLAMA-2, and Mixtral. Our findings reveal that while zero-shot learning provides convenience, it is unreliable for effective spam detection. Few-shot learning, particularly with carefully selected examples, improves detection but exhibits variability across models. Fine-tuning emerges as the most effective strategy, with Mixtral achieving 98.61\% accuracy and a balanced false positive and false negative rate below 2\%, meeting the criteria for robust spam detection. Furthermore, we explore the resilience of these models to adversarial attacks, finding that fine-tuning significantly enhances robustness against both perceptible and imperceptible manipulations. Lastly, we investigate the impact of concept drift and demonstrate that fine-tuned LLMs, especially when combined with few-shot learning, can mitigate its effects, maintaining high performance even on evolving spam datasets. This study highlights the importance of fine-tuning and tailored learning strategies to deploy LLMs effectively for real-world SMS spam detection.

\end{abstract}

\maketitle   

\section{Introduction}

The prevalence of SMS spam has increased significantly in recent years, driven in part by the rise of application-to-person (A2P) bulk messaging services such as Twilio~\footnote{https://www.twilio.com/}, which facilitate mass delivery of SMS messages to mobile users, and the increased use of SMS in online applications (social media apps, banking, etc.), including multifactor authentication and application-to-customer communications such as account alerts. In the first half of 2023, Americans received 78 billion scam messages, marking an increase of 18\% compared to the first half of 2022 \cite{robo23}. These messages resulted in estimated losses of \$13 billion, which is \$4 billion higher than the losses reported in the same period of the previous year. Similarly, the Australian Competition and Consumer Commission (ACCC)'s ScamWatch body reported a significant rise in text scams. In 2023, there were 109,615 SMS fraud reports, an increase from 79,835 in 2022. In the first half of 2024 alone, 57,533 text scams were reported by ScamWatch \cite{accs}.

Researchers have proposed various models ranging from conventional machine learning techniques to state-of-the-art deep learning and transformer-based architectures to counter the threat of SMS spam. 
However, machine learning-based models are known to be prone to adversarial perturbations. Salman et al. examined the robustness of various ML-based spam detectors—including conventional ML models, one-class and PU machine learning models, neural networks, and deep learning models using transformer-based architectures—as well as mobile text filtering apps and third-party anti-spam services, finding all of them susceptible to adversarial manipulations.
Moreover, their effectiveness is hindered by the constantly evolving nature of spam. This shift in spam tactics, known as concept drift, undermines the ability of these models to generalize and perform effectively on new unseen SMS spam \cite{salman2024investigating}. 
Several, recent studies indicate the failure of current spam detection infrastructure, including AI-enabled mobile text filtering apps and third-party anti-spam services, on recent spam datasets \cite{salman2024investigating,tang2022clues}.

Recent studies have highlighted the impressive performance of large language models (LLMs) across various natural language processing (NLP) tasks \cite{fatemi2023comparative,ge2024openagi}. Models such as GPT-4 (1.76T) \cite{achiam2023gpt}, LLaMA \cite{touvron2023llama}, and Mixtral \cite{jiang2024mixtral} have shown exceptional capabilities in a wide range of NLP applications. These advanced AI models, trained on extensive text and code datasets, have excelled in tasks ranging from natural language translation to text summarization \cite{kalyan2023survey}. LLMs have proven particularly effective for text classification tasks \cite{fields2024survey}, where the objective is to assign a predefined label to a given text. In the context of spam classification, the problem can be approached as a text classification task where the LLM processes an SMS and generates a label, such as spam or ham. Recently, Apple Inc. announced its collaboration with OpenAI to integrate GPT-4 (1.76T) into various components of iOS, iPadOS, and macOS, including Siri and writing tools \cite{openAiApple}. This development highlights the potential of using advanced NLP techniques and cutting-edge LLMs to improve SMS spam filtering.

Large language models (LLMs), trained on extensive textual datasets, are skilled at recognizing linguistic patterns, enabling them to detect SMS spam using zero-shot and few-shot learning methods without explicit training \cite{brown2020language,yang2024harnessing}. These models show promise for detecting SMS spam, a domain often lacking new spam datasets. While zero-shot and few-shot learning have demonstrated impressive outcomes across various NLP tasks, their performance occasionally falls short of fine-tuned approaches \cite{brown2020language,liu2022few}. Investigating the effectiveness of zero-shot and few-shot learning compared to fine-tuning in SMS spam detection is essential, as this comparison is underexplored in the existing literature. Additionally, incorporating ``chain of thoughts" prompting could enhance model interpretability and performance in complex scenarios, a dimension not yet extensively studied, particularly under adversarial perturbation and concept drift conditions.

To this end, our contributions are as follows.
\begin{itemize}[leftmargin=*]
    \item \descr{Evaluation of Zero-shot and Few-shot Learning, and Chain-of-thought Prompting.} We investigated the performance of commercial and open-source models with zero- and few-shot learning under different settings, including Chain-of-thought Prompting. To this end, we selected two commercial LLMs, GPT-4 (1.76T) and DeepSeek (236B), and two open-source LLMs, LLAMA-2 (70B) and Mixtral (8$\times$7B), along with their smaller counterparts, LLAMA-2 (13B) and Mistral (7B). We found that the performance of LLMs was not \textit{Satisfactory} (\S~\ref{sec:spam_catg}) under the zero-shot settings (\S~\ref{sec:zeroshot_eval}), with none achieving a balanced low false positive rate (FPR) and false negative rate (FNR). However, in the few-shot learning setting (\S~\ref{sec:fewshot_eval}), GPT-4 demonstrated \textit{Satisfactory} performance with 97.18\% accuracy, though it exhibited a slightly higher FNR of 4.97\%. Our findings indicate that system prompts show poor \textit{transferability} across different LLMs. The performance of GPT-4 was further improved with Chain-of-thought Prompting, achieving an accuracy of 97.60\% (\S~\ref{sec:cot_eval}). Contrary to recent work on the performance gains of smaller code-oriented LLMs~\cite{dunlap2024pairing}, we found that smaller chat-oriented LLMs are not robust to SMS spam; however, the larger LLMs demonstrate exceptional robustness. Additionally, our analysis highlights that smaller LLMs are more sensitive to system prompts than larger LLMs (\S~\ref{llm_size_impact}).

    \item \descr{Fine-tuning LLMs for Spam Detection.} We extend our experiments by fine-tuning LLMs on a recent spam dataset. Specifically, we focus on the three top-performing open-source LLMs such as LLAMA-2 (70B), LLAMA-2 (13 B), and Mixtral (8$\times$7B), to determine whether their performance improves over zero- and few-shot learning approaches  (\S~\ref{sec:finetunning_eval}). We found that Mixtral (8$\times$7B) achieved an accuracy of 98.61\% with a balanced FPR and FNR below 2\% meeting the desired criteria for \textit{Good} (\S~\ref{sec:spam_catg}) spam detector.

    \item \descr{Robustness to Adversarial Manipulation.} We conduct a robustness analysis of all the LLMs in zero-shot settings and with the fine-tuned LLMs to explore their resilience to adversarial attacks. In particular, we evaluated the LLMs on both perceptible and recent imperceptible perturbations(\S~\ref{sec:attacks_eval}). 
    We found that while most LLMs are inherently robust to adversarial manipulations; however, this robustness is not universal across all models.

    \item \descr{Profiling LLMs Against Concept Drift.} Finally, we profiled the performance of the best-performing LLM, Mixtral-70B, against concept drift by fine-tuning the model on an older dataset and testing it against a recent spam set. We compared its performance with several baseline models, each trained and tested similarly (\S~\ref{sec:drift_eval}). We found that while the performance of the baseline models dropped drastically, the LLMs experienced significantly less impact, demonstrating their potential as the best choice for spam detection.
\end{itemize}

Our study addresses the persistent challenges in tackling SMS spam, particularly the issues of robustness to adversarial manipulation and concept drift. Through our analysis, we aim to draw the research community's attention to the need for designing LLM-based spam detectors that are suitable for real-world applications and capable of countering the evolving tactics employed by spammers. Such advancements are crucial for better protecting the general public from the ongoing threat of SMS spam.

\section{Methods and Evaluation Framework}
\begin{figure*}[th]
\centering
\includegraphics[width=0.8\linewidth]{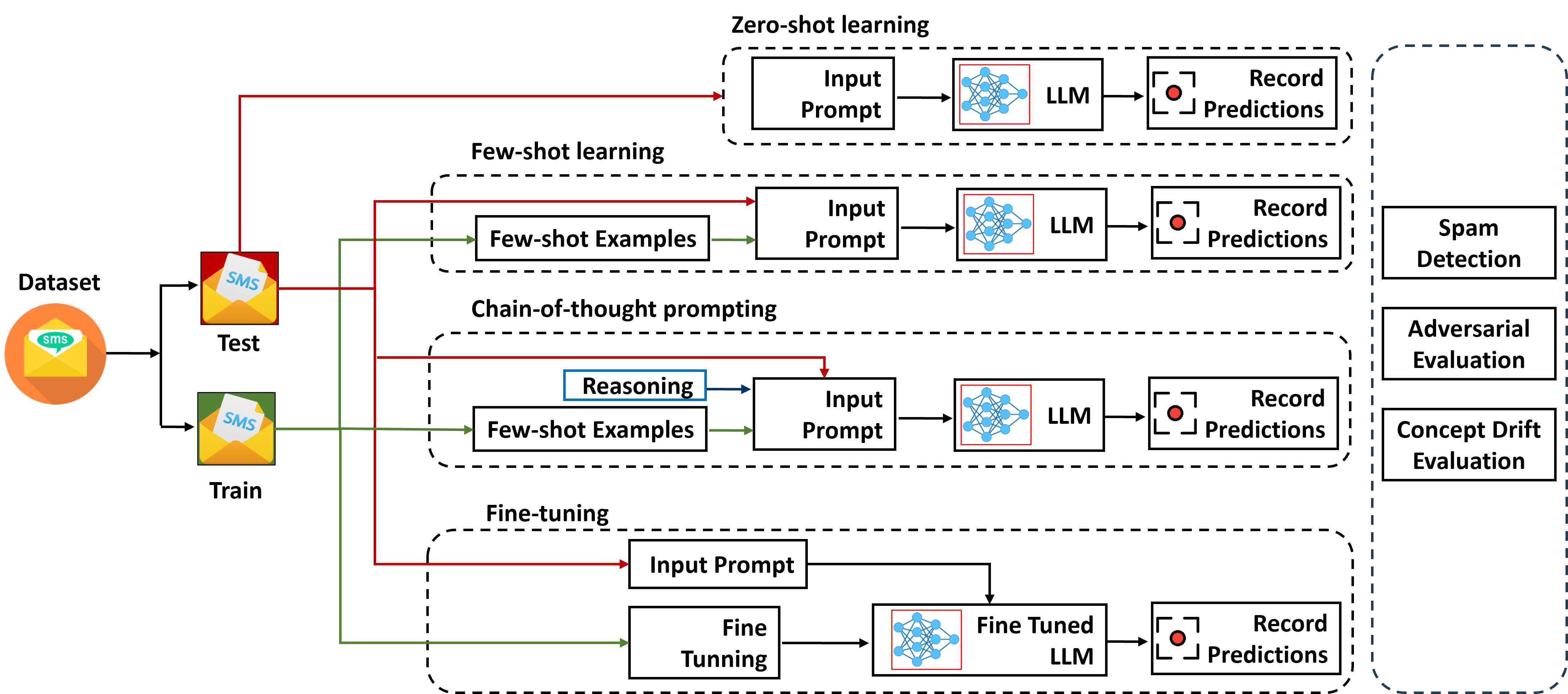}
\caption{Evaluation framework of LLMs using zero-shot, few-shot, fine-tuning, and chain-of-thought prompting approaches. The process includes training and testing on SMS datasets, with subsequent assessment through spam detection, adversarial evaluation, and concept drift evaluation.}
\label{fig:llm_eval}
\end{figure*}

In the following sections, we detail our methodologies and evaluation setup, outlining the dataset used, the training or learning strategies, and the attacks employed in our experiments. A high-level overview of the LLMs evaluation is shown in Figure \ref{fig:llm_eval}.

\subsection{Dataset}
\label{sec:dataset}
Our evaluation is based on the recent ``Super Dataset'' \cite{salman2024investigating}, which contains 67,018 labeled SMS messages collected over a decade (2012-2023). This dataset includes 40,837 benign messages (60.9\%) and 26,181 spam messages (39.1\%) from various sources. The Super Dataset aggregates SMS spam data from widely used datasets, including the UCI SMS Spam Dataset (released in 2012) \cite{almeida2013towards}, the NUS SMS Dataset (released in 2015) \cite{chen2013creating}, and the latest Spamhunter Dataset (released in 2022) \cite{tang2022clues}. Furthermore, several hundred benign and spam messages were collected from volunteers and various other sources, including Twitter and scam observatories such as the ScamWatch \cite{accs} and the UK Action Fraud \cite{actionfraud}.

We split the Super Dataset into training and test sets. The test set was created by randomly selecting 5,000 benign and 5,000 spam SMS messages, while the remaining SMS messages were used for the training set. Additionally, we created a ``holdout set'' by randomly selecting 300 spam messages from the test set. This holdout set was used to apply transformations for generating adversarial examples.

\subsection{LLMs Learning Methods}
\subsubsection{Zero-shot and Few-shot Learning}
\label{sec:zero-few}
We evaluate proprietary LLMs such as GPT-4 (1.76T) \cite{gpt4} and DeepSeek (deepseek-chat 236B) \cite{deepseek}, as well as state-of-the-art open-source LLMs such as LLAMA-2\_70B (Llama-2-70b-chat-hf) \cite{llama70B} and Mixtral\_8$\times$7B (Mixtral-8$\times$7B-Instruct-v0.1) \cite{mixtral}. We also consider their smaller counterparts, including LLAMA-2\_13B (Llama-2-13b-chat-hf) \cite{llama13B} and Mistral\_7B (Mistral-7B-Instruct-v0.3) \cite{mistral}. We investigate the zero-shot and few-shot learning capabilities of these LLMs by assessing their performance on SMS spam detection without any specific training or modification of the model weights or parameters.

\begin{itemize}[leftmargin=*]

\item \descr{Zero-shot learning} is a technique where the model is given only the task name, a description of the task, and the label space—comprising a set of possible options guiding the model's response. Unlike methods that include demonstrations, zero-shot learning provides just a task description and a prompt. This method is the most convenient, robust, and challenging, as no demonstrations are supplied. This approach enables models to tackle tasks they were not specifically trained on, leveraging their ability to generalize from related tasks.

 \item \descr{Few-shot learning} involves providing the model with task definitions and a small number of input-output pairs, consisting of a few randomly selected labeled examples for each class. This approach trains models with a small number of examples for each class, giving them just enough data to perform the task effectively. However, selecting the appropriate examples to include in the prompt is critical for maximizing the effectiveness of few-shot learning. We assessed the few-shot learning capabilities of LLMs in various scenarios and with different sets of examples, demonstrating their ability to generalize from limited examples and perform spam classification with minimal supervision. This technique is especially valuable for tasks where labeled data is scarce or costly to obtain.

\end{itemize}

\subsubsection{Chain-of-Thought Prompting \cite{wei2022chain}.} This approach guides the model through a series of intermediate reasoning steps, simulating a logical thought process. This method improves the model's ability to perform complicated tasks by breaking them down into smaller, more manageable parts. We evaluated the effectiveness of Chain-of-Thought Prompting with LLMs in the context of SMS spam detection, illustrating their capacity to follow and generate coherent, step-by-step reasoning. This approach is particularly beneficial for tasks that require multi-step analysis, such as distinguishing between spam and legitimate messages, offering improved performance and interpretability even with limited direct supervision. Although it can be used within Few-Shot Learning, Chain-of-Thought Prompting primarily focuses on the structured approach to reasoning rather than the amount of data used (see Appendix \ref{sec:cot_appendix} for details).

\subsubsection{Fine-tuning}
\label{sec:finetune}
Fine-tuning pre-trained LLMs for specific tasks usually requires significant computational resources due to the need to update all model parameters. To mitigate this, Microsoft researchers introduced Low-Rank Adaptation (LoRA) \cite{hu2021lora}, which adds sets of rank-decomposed weight matrices to existing model weights. During fine-tuning, only the newly added weights are updated, while the pre-trained weights remain unchanged. This method is further improved by QLoRA \cite{dettmers2024qlora}, which backpropagates gradients via a 4-bit quantized pre-trained model into the LoRA, making the process more memory-efficient and computationally faster while achieving superior performance compared to other fine-tuning methods.

In our study, we selected Meta's LLAMA-2\_70B and Mixtral (8$\times$7B) for finetunning. LLAMA 2 is an open-source fine-tuned generative text model optimized for dialogue use cases. The Mixtral (8$\times$7B) LLM is a pre-trained generative Sparse Mixture of Experts, consisting of multiple sub-models called ``experts'', with Mixtral (8$\times$7B) having 8 experts, each specialized to optimize decisions for different tasks.

We deployed LLAMA2 and Mixtral in our local GPU environment to assess their effectiveness in SMS spam detection on original SMS texts and in the presence of adversarial examples.

\subsection{Threat Model}
We assume a black-box scenario in which the adversary has no direct access to the internal details of the model, such as the architecture, weights, or parameters. Instead, the adversary can only interact with the model by providing inputs and observing the corresponding outputs (predictions). Moreover, similar to the domain of image classification, we include bounds or constraints to adversarial perturbations to make the adversarial examples indistinguishable from the benign original. The adversary aims to modify an SMS so that it is misclassified by the target spam detector while retaining semantic meaning and imperceptibility.

\subsection{Adversarial Attacks}
\label{sec:attacks}
After a thorough review of recent literature on black-box attacks, we include the following attacks in our evaluation.

\descr{Perceptible.} To retain the same semantic meaning and near imperceptibility while modifying the text, we established the following constraints:

\begin{itemize}[leftmargin=*]
    \item \textbf{Thesaurus Creation}: We created a thesaurus of the 500 most common keywords found in the spam messages within the dataset, utilizing the Python library WordCloud \cite{heimerl2014word}. Only spam keywords from the thesaurus may be manipulated.
    \item \textbf{Single Character Perturbation}: We constrained the modification of spam keywords to a single character in each spam keyword to limit the perturbation.
    \item \textbf{Preservation of URLs and Contact Details}: We avoided altering URLs and contact details, such as email addresses and phone numbers.
\end{itemize}

To this end, we implemented six black-box adversarial tactics listed in Table \ref{tab:attacks}. We extended the TextAttack framework\footnote{TextAttack is a Python library designed for conducting adversarial attacks, enhancing data through augmentation, and training models within the field of natural language processing (NLP).} \cite{morris2020textattack} to allow for the modification of constraints for generating adversarial examples (see Table \ref{tab:attacks}(b-f)). Specifically, we modified TextAttack to incorporate our constraints on thesaurus-based keyword manipulations, single-character perturbations, and preservation of URLs and contact details. Additionally, although TextAttack does not natively support spacing attacks, we extended its codebase to implement this feature, enabling the generation of spacing adversarial examples (see Table \ref{tab:attacks} (a)). This enhancement allowed us to evaluate the impact of spacing manipulations on the robustness of our SMS spam detection models.

\descr{Imperceptible.} A recent black-box text attack introduced by Boucher et al. \cite{boucher2022bad} has shown remarkable efficacy against a variety of commercially deployed Natural Language Processing (NLP) systems from companies such as Microsoft and Google, as well as open-source models by Facebook, IBM, and HuggingFace. This attack stands out from prior methods due to the subtlety of the adversarial examples it generates. The attack introduces imperceptible changes to the original text, which preserves the stealthiness while effectively deceiving the target models. This is accomplished by inserting non-printable Unicode characters or replacing characters with visually similar homoglyphs. We adopted these imperceptible attacks using their open-source code \cite{imperceptible_perturb}, as detailed in Table  \ref{tab:attacks}.

\subsection{Performance Categorization}
\label{sec:spam_catg}

While there is no agreement over the acceptable False Positive Rate (FPR) and False Negative Rate (FNR) for SMS spam detection in the literature, we define three distinct levels to evaluate the performance of LLMs in this study: \textit{Satisfactory (FPR \& FNR < 5\%)}, \textit{Good (FPR \& FNR < 3\%)}, and \textit{Perfect (FPR \& FNR < 1\%)}. These categories help in understanding the trade-offs between user experience and the LLM's ability to accurately detect spam (see Appendix \ref{sec:spam_catg_appendix}).

\subsection{System Prompt}
LLMs depend on system prompts to shape their actions and outputs. We designed specific prompts for zero-shot, few-shot, and Chain-of-Thought settings, following the best practices provided by OpenAI \cite{prompteng} to instruct the LLM in determining whether an SMS should be labeled as spam or ham. All three prompts direct the model to respond with either 'Spam' or 'Ham.' The primary distinction between the zero-shot and few-shot prompts is that the few-shot prompt includes example data to provide context and guidance. Additionally, the Chain-of-Thought prompt goes further by incorporating extensive reasoning steps that the model should consider before classifying a message as spam or ham. The prompts for zero-shot, few-shot, and Chain-of-Thought are shown in Figures \ref{fig:zero_shot_prompt}, \ref{fig:few_shot_prompt}, and \ref{fig:cot_prompt} in the Appendix, respectively.

\section{Evaluation}

\subsection{Evaluation of LLMs with Zero-shot Learning}
\label{sec:zeroshot_eval}

In this section, we evaluate the performance of different pre-trained LLMs in detecting SMS spam using the test set (see \ref{sec:dataset}). The assessment includes models of different scales, specifically comparing smaller and larger versions of the LLAMA and Mixtral models. In addition, we include commercial models such as DeepSeek (236B) and GPT-4 (1.76T). The evaluation is conducted in a zero-shot learning context. The primary objective is to determine how effectively each LLM can identify spam messages without prior exposure to specific examples, leveraging their pre-trained capabilities. Please refer to listing \ref{fig:zero_shot_prompt} (in appendix) for the zero-shot prompt.   The research posits that larger models, due to their increased capacity and training data, will outperform smaller models in terms of accuracy and robustness in spam detection tasks.

Table \ref{tab:zeroshot} summarizes the performance of each model in a zero-shot learning setup. Larger models generally outperformed their smaller counterparts, with Mixtral (8$\times$7B) and GPT-4 (1.76T) showing particularly high accuracy. Notably, LLAMA (70B) demonstrated a high TPR of 98.36\% but suffered from a low TNR of 47.44\%, indicating a tendency to mislabel legitimate messages as spam. In contrast, GPT-4 exhibited exceptional TNR at 99.80\%, showcasing its ability to accurately identify legitimate messages, though this came at the cost of a higher FNR of 14.31\%. Mixtral (8$\times$7B) achieved a balanced performance, with an accuracy of 87.43\%, a TPR of 91.95\%, and a TNR of 83.18\%, making it one of the most reliable models in this evaluation. Meanwhile, DeepSeek (236B) also performed well, with a high accuracy of 91.90\% and a strong TNR of 99.20\%, though it struggled with a higher FNR of 15.42\%. In contrast, smaller models like LLAMA (13B) and Mistral (7B) showed more variability and generally lower performance. LLAMA (13B) had difficulty balancing TPR and TNR, with both metrics hovering around 52\%, while Mistral (7B) performed moderately well with an accuracy of 80.40\%, but it also displayed a relatively high FPR of 36.56\%. Despite the promising results from larger models, none of the LLMs achieved a \textit{satisfactory} balance of low FPR and FNR simultaneously, indicating that their performance in zero-shot spam detection is not yet reliable.

\begin{table}[hbt!]
\renewcommand{\arraystretch}{1.15}
\tabcolsep=0.05cm
\begin{center}
\caption{Performance evaluation of {\it LLMs} with Zero Shot Learning. Larger models like Mixtral (70B) and GPT-4 outperform smaller models, but none achieve a \textit{satisfactory} balance between low FPR and FNR.} 
\label{tab:zeroshot}
\begin{tabular}{ c | c | c | c | c | c | c }
\toprule
 \hline
 & \multicolumn{6}{c}{\bf Performance Metrics}\\
\cline{2-7}
 \textbf{Model} & \textbf{Acc} & \textbf{FS} & \textbf{TPR} & \textbf{TNR} & \textbf{FPR} & \textbf{FNR} \\ [0.5ex]
 \midrule\hline
LLAMA (70B) & 72.13\% & 77.38\% & 98.36\% & 47.44\% & 52.56\% & 1.64\% \\
LLAMA (13B) & 55.14\% & 43.06\% & 52.43\% & 52.43\% & 47.57\% & 47.57\% \\
\textbf{Mixtral (8$\times$7B)} & \textbf{87.43\%} & \textbf{87.64\%} & \textbf{91.95\%} & \textbf{83.18\%} & \textbf{16.82\%} & \textbf{8.05\%} \\
Mistral (7B) & 80.40\% & 83.23\% & 97.40\% & 63.44\% & 36.56\% & 2.60\% \\
DeepSeek (236B) & 91.90\% & 91.25\% & 84.58\% & 99.20\% & 0.80\% & 15.42\% \\
\textbf{GPT-4 (1.76T)} & \textbf{92.75\%} & \textbf{92.19\%} & \textbf{85.69\%} & \textbf{99.80\%} & \textbf{0.20\%} & \textbf{14.31\%} \\
\hline
  \bottomrule
\end{tabular}
\end{center}
\end{table}

\subsection{Evaluation of LLMs with Few-shot Learning}
\label{sec:fewshot_eval}
Next, we evaluate the performance of LLMs in a few-shot learning context for detecting SMS spam using the test set (see \ref{sec:dataset}). The objective is to determine how effectively each LLM can identify spam messages when provided with a few-shot learning prompt that includes a set of examples. These examples guide the LLM through the context and particulars of the task. To ensure a balanced set of examples, we selected ten spam messages, five from each of the two categories of spam i.e. Fraud\footnote{A fraudulent message attempts to mislead SMS recipients into taking actions like clicking a URL or calling a specified number.} and Promotional (Promo)\footnote{A promotional message seeks to advertise goods or services or to disseminate opinions.}\cite{tang2022clues}.

Table \ref{tab:perform-few-shot} summarizes the performance metrics for LLMs under few-shot learning. While few-shot learning is generally assumed to improve model performance, the results in Table \ref{tab:perform-few-shot} indicate that this is not uniformly the case. The performance of various LLMs when tested with few-shot learning examples reveals a more complex picture compared to their zero-shot learning performance.
The few-shot prompt design can be seen in Figure \ref{fig:few_shot_prompt} in the appendix.

\textbf{LLAMA (70B)}, for instance, shows mixed results. In the few-shot setup, its accuracy improved slightly to 79.37\% from 72.13\% in the zero-shot scenario. However, this improvement came alongside a significant drop in its TNR, from 98.36\% to 91.29\%. This suggests that while the model became somewhat better at detecting benign (as indicated by the comparatively higher TNR of 68.16\%, than the 47.44\% in the zero-shot scenario), it struggled more with identifying spam messages, leading to a higher rate of false negatives. On the other hand, \textbf{Mixtral (8$\times$7B)}, which performed well in the zero-shot setup with an accuracy of 87.43\% and a TPR of 91.95\%, saw its accuracy drop to 74.18\% in the few-shot setup. The TNR also fell from 83.18\% to 54.86\%, and the FPR increased from 16.82\% to 45.14\%, indicating a degradation in performance when the model was exposed to diverse few-shot examples.

A similar trend is observed for \textbf{DeepSeek (236B)}, where its accuracy decreased from 91.90\% in the zero-shot context to 83.35\% in the few-shot context. While its TNR increased slightly from 99.20\% to 99.50\%, the TPR dropped significantly from 84.58\% to 67.17\%, showing that the model became less effective at identifying spam messages when given few-shot examples. \textbf{GPT-4 (1.76T)}, which achieved the highest zero-shot accuracy at 92.75\%, also experienced a decline in performance, with its accuracy dropping to 87.40\% in the few-shot scenario. Its TPR decreased from 85.69\% to 74.87\%, and while its TNR remained nearly perfect at 99.90\%, the overall performance suggests that few-shot learning did not provide the anticipated improvement.

Overall, these results indicate that few-shot learning does not consistently enhance model performance and, in many cases, leads to a degradation in the model's ability to correctly identify both spam and legitimate messages. Moreover, the variations in performance across different LLMs suggest that a prompt performing well on one LLM does not guarantee good results on another, highlighting the poor transferability of developed prompts across different models. This underscores the importance of carefully selecting and tailoring examples for each LLM in few-shot learning, as a ``one-size-fits-all'' approach is ineffective across different LLMs. Each model may require prompts specifically aligned with its unique strengths and weaknesses to achieve optimal performance in tasks like spam detection.

\begin{table}[hbt!]
\renewcommand{\arraystretch}{1.15}
\tabcolsep=0.05cm
\begin{center}
\caption{Baseline performance evaluation of LLMs with few-shot learning using randomly selected, balanced examples from both categories of spam. Few-shot learning enhances performance for some LLMs but leads to inconsistent results across different models.}
\label{tab:perform-few-shot}
\begin{tabular}{ c | c | c | c | c | c | c }
\toprule
 & \multicolumn{6}{c}{\bf Performance Metrics}\\
\cline{2-7}
 \textbf{Model} & \textbf{Acc} & \textbf{FS} & \textbf{TPR} & \textbf{TNR} & \textbf{FPR} & \textbf{FNR} \\ [0.5ex]
 \midrule
\textbf{LLAMA (70B)} & \textbf{79.37\%} & \textbf{81.10\%} & \textbf{91.29\%} & \textbf{68.16\%} & \textbf{31.84\%} & \textbf{8.71\%} \\
LLAMA (13B) & 50.95\% & 67.07\% & 100.00\% & 2.00\% & 98.00\% & 0.00\% \\
Mixtral (8$\times$7B) & 74.18\% & 83.23\% & 95.35\% & 54.86\% & 45.14\% & 4.65\% \\
Mistral (7B) & 52.35\% & 66.69\% & 100.00\% & 6.12\% & 93.88\% & 0.00\% \\
DeepSeek (236B) & 83.35\% & 80.12\% & 67.17\% & 99.50\% & 0.50\% & 32.83\% \\
\textbf{GPT-4 (1.76T)} & \textbf{87.40\%} & \textbf{85.58\%} & \textbf{74.87\%} & \textbf{99.90\%} & \textbf{0.10\%} & \textbf{25.13\%} \\
  \hline
  \bottomrule
\end{tabular}
\end{center}
\end{table}

\subsection{Ablation Analysis}

To explore the sensitivity of LLMs to the number, type, and length of spam exemplars in few-shot learning, we analyzed the classification outcomes of each LLM to identify where they were struggling. Based on the findings from this misclassification analysis, we designed several experiments aimed at assessing how the choice of examples and their characteristics affect the models' performance in detecting SMS spam.

\subsubsection{Analysis of Misclassification of Zero-Shot Learning}
In evaluating LLMs using zero-shot and few-shot learning approaches, neither method achieved a balanced FPR and FNR. While few-shot learning improved the performance of some LLMs, it downgraded the performance of others. This leads us to believe that some LLMs are sensitive to certain types of SMS messages, while others respond differently. Therefore, to understand and identify the sensitivity of each LLM to various types of SMS messages, we performed this analysis on zero-shot learning to help us select more tailored few-shot examples for improved spam detection.

\descr{LLAMA-2\_70B (Llama-2-70b-chat-hf).}
In the context of zero-shot learning, the common reasons for \textit{False positives} by LLAMA 70B include the presence of keywords typically associated with scams, such as ``offer'', ``win'', ``prize'', ``free'', and ``account''. URLs and phone numbers also contribute to false positives due to their frequent appearance in both scam and benign messages. Elements of social engineering that imply urgency or authority often lead to misclassifications. For example, a benign message like ``Hmm I think shld be ok? I confirm with you again later.'' and ``YES! We should all have dinner! I miss you dearly!'', which were flagged due to keywords and enthusiastic tone, respectively. Similarly, common reasons for \textit{false negatives} include messages that lack explicit scam keywords or contain subtle social engineering signals. Shorter messages are particularly challenging for the model, as they may not provide sufficient context for accurate classification. For instance, a message like ``Chase Bank Password Reset Code XXXXXX If you did not request this, please call us immediately.'' was not flagged as a scam, despite containing clear indicators such as an urgent call to action and reference to a financial institution.

\descr{LLAMA-2\_13B (Llama-2-13b-chat-hf).} The model is optimized for both efficiency and effectiveness in security-related content analysis, making it highly suitable for detecting urgency and threats within messages. It is robust in scenarios involving security alerts or urgent requests, accurately classifying messages like ``Security Alert: Unusual activity detected in your account. Click here to secure it.'' However,  the model's focus on security-related content might lead to oversensitivity to any message mentioning security, regardless of its actual intent. For instance, benign reminders about account security, such as updates or regular checks, might be incorrectly flagged as potential scams. Additionally, it faces difficulties with non-standard language uses, which can lead to misclassifications, especially in less formal communications. A benign reminder SMS like ``Don't forget to update your password regularly for security,'' might be misinterpreted as a scam due to its emphasis on security.

\descr{Mixtral (8$\times$7B) Instruct-v0.1} 
The misclassification analysis suggests that Mixtral adopts a balanced approach, weighing certain keywords heavily, which aids in detecting scams but can also lead to false positives. The model's performance is influenced significantly by the content of the messages, and it can struggle with informal language and slang common in SMS communications. In case of False positives, some benign messages are misclassified as scams, likely due to the presence of trigger words commonly found in scam messages. An example of a benign message that might be incorrectly flagged is one that discusses a ``transaction'' or ``confirmation''.  
Similarly, in the case of False negatives, the model misses some subtle scam messages, particularly those lacking overt scam indicators or closely mimicking benign message formats. An example of this could be a scam disguised within a legitimate-looking service update or notification.

\descr{Mistral (8$\times$7B) Instruct-v0.3.} 
While smaller than Mixtral (8$\times$7B), is designed for efficient processing with a keen focus on detecting thematic elements. This model integrates well with datasets that involve specific themes, such as finance and promotions. It is particularly effective at identifying scams that use promotional language or financial incentives, Mistral adeptly flags messages like ``Flash Sale: Credit cards accepted with 0\% interest for the first year!''. Nevertheless, its smaller size and thematic focus may limit its ability to process and understand less straightforward, mixed-language, or jargon-heavy messages. Technical service messages, especially those with mixed language usage, often elude Mistral's detection capabilities, reflecting a potential area for further model training. While Mistral is adept at theme recognition, its challenges with linguistic diversity suggest a need for more inclusive language modeling.

\descr{Deepseek V2 Chat (deepseek-chat 236B).} The model excels in identifying urgent and direct calls to action, and uses a streamlined architecture that allows it to process such messages quickly and effectively. Its ability to detect and flag urgent calls to action makes it excellent for identifying clear phishing attempts, such as ``Immediate action required: Verify your account to prevent closure''. However, its streamlined focus may miss nuances in more complex or subtly worded messages, leading to possible false negatives. For instance, simple directives or appointments like ``Confirm your appointment by calling this number'' may be misinterpreted as scams due to their direct and urgent wording. It is further limited in processing sparse text or niche vocabulary, often misinterpreting benign intentions. An SMS like "Meet me at the cafe at 3 PM" could be classified incorrectly due to its concise and direct nature, which may mimic the brevity of scam messages.

\descr{GPT-4 (1.76T).} As one of the largest models in this analysis, GPT-4 boasts extensive training data and architectural complexity, which provide it with a broad comprehension of linguistic nuances and the ability to generalize across various text types. Known for its deep contextual understanding, GPT-4 excels in identifying clear scam indicators and nuanced language patterns within SMS. It can effectively contextualize and classify straightforward scam messages such as ``Win \$1000 now by clicking on this link!''. Despite its size and capabilities, GPT-4 sometimes struggles with highly colloquial text or messages that use niche-specific jargon, which might not be sufficiently represented in its training data. A message like ``Yo, hit up this link to grab your free tix'' might be misclassified due to the casual slang and ambiguous context, potentially leading to a false negative in scam detection. GPT's broad linguistic comprehension makes it valuable for diverse scenarios but highlights the need for additional training to handle complex linguistic nuances better.

This analysis reveals distinct strengths and weaknesses in each model, all of which are crucial for effective scam detection. The performance differences among these models stem largely from their unique training processes and architectural characteristics. While GPT-4 and Deepseek are comprehensive, they are vulnerable to subtle linguistic variations, indicating potential training gaps. However, Mixtral and LLAMA, despite their thematic precision, struggle with linguistic diversity, highlighting the need for more inclusive language modeling. All models, however, encounter difficulties with complex linguistic features such as jargon, slang, or mixed languages, leading to higher misclassification rates. In addition, benign messages that resemble the format of scam messages often result in misclassifications. This analysis emphasizes the importance of selecting an optimal set of examples to enhance the spam detection capabilities of LLMs in few-shot learning.

\subsubsection{Sensitivity of LLMs to Spam Exemplars in Few-Shot Learning}

\descr{Few-Shot Learning with Varying Number of Spam Exemplars.}
For all the LLMs, we began our evaluation with few-shot learning using 5 examples, consisting of two examples each from both categories of spam (i.e., Fraud and Promo) and one benign example. The LLMs were then evaluated on a smaller set of 400 SMS messages (200 spam, 200 ham). The examples for each LLM were selected based on the misclassification analysis, identifying the specific types of SMS messages each LLM struggled with. The experiment was repeated multiple times for each LLM with different SMS messages containing similar keywords until the highest accuracy was no longer increased. The results are provided in Table \ref{tab:perform-message-number}. Similar experiments were conducted for few-shot learning with 10 examples (4 Fraud, 4 Promo, 2 ham), few-shot learning with 15 examples (6 Fraud, 6 Promo, 3 ham), and few-shot learning with 20 examples (8 Fraud, 8 Promo, 4 ham) to explore how the number of spam examples provided during few-shot learning influences the performance of LLMs in detecting SMS spam. By varying the number of examples, we aimed to determine the sensitivity of LLMs to the amount of training data and to identify the optimal number of examples that yield the best performance in different metrics. We hypothesized that increasing the number of examples would improve the models' ability to accurately classify spam messages, particularly in terms of balancing true positive and true negative rates.

From Table \ref{tab:perform-message-number}, it can be seen that with 5 examples, all of the LLMs' accuracy improved over the previous few-shot learning results (see Table \ref{tab:perform-few-shot}). Moreover, with the exception of Mistral, all other models not only outperformed the few-shot learning but also the zero-shot approaches. LLAMA (70B) achieved 89.01\% accuracy and a balanced TPR and TNR, indicating robust performance with a moderate number of examples. LLAMA (13B) showed a strong TPR (97.30\%) but a lower TNR (73.57\%), suggesting a tendency towards higher false positives. DeepSeek (236B) excelled with 95.10\% accuracy, high TPR (91.09\%), and TNR (99.10\%). GPT-4 (1.76T) also demonstrated high accuracy (93.95\%) and balanced metrics, indicating effective adaptation to the example set.

When the number of examples was increased to 10, the performance of most LLMs showed a downward trend compared to their performance with 5-shot examples, except for GPT-4. \textbf{GPT-4} achieved the highest accuracy of 96.35\% with 10-shot examples, showing an improvement over its already strong performance of 93.95\% with 5-shot examples. This suggests that GPT-4 benefits from the increased number of examples, particularly in reducing false negatives, as indicated by its FNR, which decreased from 12.01\% to 7.11\%. These results suggest that while increasing the number of examples to 10 can improve the performance of certain models like GPT-4, the impact is not universally positive across all models, with many experiencing marginal declines in their performance. 
Similarly, with 15-shot examples, the performance of most LLMs further declined compared to both the 5-shot and 10-shot scenarios, except for GPT-4, which experienced a decrease in accuracy compared to the 10-shot but remained higher than with 5-shot examples.

In conclusion, our experiments reveal that few-shot learning improves the performance of most LLMs in detecting SMS spam with tailored examples, however, the optimal number of examples varies by model. While some models benefit significantly from additional examples, others do not show the same level of improvement. These findings highlight the importance of tailoring the number and type of few-shot examples to each specific LLM to achieve the best balance between true positive and true negative rates.

\begin{table}[hbt!]
\renewcommand{\arraystretch}{1.15}
\tabcolsep=0.05cm
\begin{center}
\caption{Performance evaluation of LLMs with few-shot learning using varying numbers of spam exemplars. Few-shot learning improves the performance of most LLMs in detecting SMS spam with tailored examples, increasing the number of examples does not consistently lead to better results.} 
\label{tab:perform-message-number}
\scalebox{0.95}{%
\begin{tabular}{ c | c | c | c | c | c | c }
 \hline
 & \multicolumn{6}{c}{\bf Performance Metrics}\\
\cline{2-7}
 \textbf{Model} & \textbf{Acc} & \textbf{FS} & \textbf{TPR} & \textbf{TNR} & \textbf{FPR} & \textbf{FNR} \\ [0.5ex]
 \hline
 \multicolumn{7}{c}{Few-Shot Learning with 5 examples} \\
 \hline
\textbf{LLAMA (70B)} & \textbf{89.01\%} & \textbf{88.96\%} & \textbf{88.06\%} & \textbf{89.96\%} & \textbf{10.04\%} & \textbf{11.94\%} \\
LLAMA (13B) & 85.57\% & 87.21\% & 97.30\% & 73.57\% & 26.43\% & 2.70\% \\
Mixtral (8$\times$7B) & 87.55\% & 87.59\% & 87.99\% & 87.11\% & 12.89\% & 12.01\% \\
Mixtral (7B) & 73.25\% & 78.80\% & 99.00\% & 47.27\% & 52.73\% & 1.00\% \\
\textbf{DeepSeek (236B)} & \textbf{95.10\%} & \textbf{94.89\%} & \textbf{91.09\%} & \textbf{99.10\%} & \textbf{0.90\%} & \textbf{8.91\%} \\
GPT-4 (1.76T) & 93.95\% & 93.56\% & 87.99\% & 99.90\% & 0.10\% & 12.01\% \\
  \hline
 \multicolumn{7}{c}{Few-Shot Learning with 10 examples} \\
 \hline
\textbf{LLAMA (70B)} & \textbf{87.88\%} & \textbf{88.02\%} & \textbf{88.38\%} & \textbf{87.37\%} & \textbf{12.63\%} & \textbf{11.62\%} \\
LLAMA (13B) & 78.02\% & 82.06\% & 98.49\% & 56.69\% & 43.31\% & 1.51\% \\
Mixtral (8$\times$7B) & 83.39\% & 82.51\% & 77.44\% & 89.48\% & 10.52\% & 22.56\% \\
Mixtral (7B) & 71.74\% & 77.88\% & 99.20\% & 44.11\% & 55.89\% & 0.80\% \\
DeepSeek (236B) & 94.70\% & 94.45\% & 90.29\% & 99.10\% & 0.90\% & 9.71\% \\
\textbf{GPT-4 (1.76T)} & \textbf{96.35\%} & \textbf{96.22\%} & \textbf{92.89\%} & \textbf{99.80\%} & \textbf{0.20\%} & \textbf{7.11\%} \\
  \hline
   \multicolumn{7}{c}{Few-Shot Learning with 15 examples} \\
 \hline
\textbf{LLAMA (70B)} & \textbf{82.96\%} & \textbf{83.12\%} & \textbf{83.17\%} & \textbf{82.76\%} & \textbf{17.24\%} & \textbf{16.83\%} \\
LLAMA (13B) & 72.90\% & 78.81\% & 97.99\% & 46.33\% & 53.67\% & 2.01\% \\
Mixtral (8$\times$7B) & 80.50\% & 79.19\% & 74.15\% & 86.86\% & 13.14\% & 25.85\% \\
Mixtral (7B) & 67.42\% & 67.89\% & 69.76\% & 65.34\% & 34.66\% & 30.24\% \\
DeepSeek (236B) & 93.24\% & 93.01\% & 91.67\% & 96.42\% & 2.58\% & 8.33\% \\
\textbf{GPT-4 (1.76T)} & \textbf{95.30\%} & \textbf{95.07\%} & \textbf{90.79\%} & \textbf{99.80\%} & \textbf{0.20\%} & \textbf{9.21\%} \\

  \hline
\end{tabular}}
\end{center}
\end{table}

\descr{Few-Shot Learning with Fraud and Promo Examples.} We then evaluated the models by providing few-shot learning examples from two distinct categories of spam: fraud and promotional (promo) messages. The objective was to observe how well the LLMs adapted when given examples from only a specific category of spam and whether it could further improve the performance of the LLMs in few-shot learning. The results are shown in Table \ref{tab:perform-ml-fraud-promo}.

When provided with 5-shot Fraud examples, the accuracy of all the open-source models decreased not only compared to the few-shot learning results in Table \ref{tab:perform-few-shot} and Table \ref{tab:perform-message-number}, but also compared to the zero-shot learning results in Table \ref{tab:zeroshot}. Among the open-source models, \textbf{LLAMA (70B)} achieved the highest performance, with an accuracy of 75.80\% but a struggling TNR of 61.57\%. In contrast, the commercial models showed varying results. \textbf{DeepSeek} achieved an accuracy of 85.95\% but struggled with a low TPR (72.47\%), indicating a bias toward false negatives. This is significantly lower than its accuracy of 95.10\% in few-shot learning in Table \ref{tab:perform-message-number} and also lower than its 91.90\% accuracy in zero-shot learning in Table \ref{tab:zeroshot}. \textbf{GPT-4}, on the other hand, demonstrated relatively stable performance, achieving an accuracy of 93.05\%, which is higher than its zero-shot (92.75\%) and few-shot learning (87.40\%) results in Table \ref{tab:perform-few-shot}, but slightly lower than its few-shot learning performance (93.95\%) in Table \ref{tab:perform-message-number}.

In contrast to Fraud-based few-shot learning, few-shot learning with promotional (promo) examples demonstrated improved performance, with the accuracy of most models increasing over their zero-shot counterparts (except for Mistral) as shown in Table \ref{tab:zeroshot} and few-shot models in Table \ref{tab:perform-few-shot}. However, while the models showed better performance compared to zero-shot learning and few-shot learning with random examples, they did not surpass the accuracy achieved by the few-shot models in Table \ref{tab:perform-message-number}. This underscores the critical importance of using representative and tailored examples in few-shot learning to achieve optimal results.

\begin{table}[hbt!]
\renewcommand{\arraystretch}{1.15}
\tabcolsep=0.05cm
\begin{center}
\caption{Performance evaluation of {\it LLMs} with few-shot learning using examples from Fraud and Promo categories. Few-shot learning with promotional examples improves model accuracy over zero-shot and random examples; however, none surpasses the accuracy achieved with representative and tailored examples.}
\label{tab:perform-ml-fraud-promo}
\scalebox{0.9}{%
\begin{tabular}{ c | c | c | c | c | c | c }
 \hline
 & \multicolumn{6}{c}{\bf Performance Metrics}\\
\cline{2-7}
 \textbf{Model} & \textbf{Acc} & \textbf{FS} & \textbf{TPR} & \textbf{TNR} & \textbf{FPR} & \textbf{FNR} \\ [0.5ex]
 \hline\hline
 \multicolumn{7}{c}{Few-Shot Learning with Fraud examples} \\
 \hline
LLAMA (70B) &  71.81\% & 69.31\% & 61.57\% & 82.78\% & 17.22\% & 38.43\% \\
LLAMA (13B) & 44.65\% & 61.27\% & 100.00\% & 1.57\% & 98.43\% & 0.00\% \\
\textbf{Mixtral (8$\times$7B)} & \textbf{75.80\%} & \textbf{78.79\%} & \textbf{89.99\%} & \textbf{61.64\%} & \textbf{38.36\%} & \textbf{10.01\%} \\
Mistral (7B) & 64.23\% & 64.15\% & 63.78\% & 35.45\% & 64.55\% & 28.34\% \\
DeepSeek (236B) & 85.95\% & 83.75\% & 72.47\% & 99.40\% & 0.60\% & 27.53\% \\
\textbf{GPT-4} & \textbf{93.05\%} & \textbf{92.79\%} & \textbf{88.18\%} & \textbf{98.90\%} & \textbf{1.10\%} & \textbf{11.82\%} \\
  \hline
 \multicolumn{7}{c}{Few-Shot Learning with Promo examples} \\
 \hline
\textbf{LLAMA (70B)} & \textbf{86.20\%} & \textbf{85.99\%} & \textbf{84.05\%} & \textbf{88.38\%} & \textbf{11.62\%} & \textbf{15.95\%} \\
LLAMA (13B) & 61.93\% & 69.46\% & 99.24\% & 33.07\% & 66.93\% & 0.76\% \\
Mixtral (8$\times$7B) & 83.97\% & 82.77\% & 76.86\% & 91.11\% & 8.89\% & 23.14\% \\
Mistral (7B) & 71.58\% & 71.23\% & 70.92\% & 34.17\% & 65.83\% & 29.08\% \\
\textbf{DeepSeek (236B)} & \textbf{93.56\%} & \textbf{93.25\%} & \textbf{85.00\%} & \textbf{99.02\%} & \textbf{0.98\%} & \textbf{15.00\%} \\
GPT-4 & 93.55\% & 93.32\% & 89.99\% & 99.50\% & 0.50\% & 10.01\% \\
  \hline
\end{tabular}}
\end{center}
\end{table}

\descr{Few-Shot Learning with Varying Message Lengths.} We further examined the impact of message length by categorizing examples into short (tokens < 20), medium (20 $\leq$ tokens < 30), and lengthy (tokens $\geq$ 30) messages based on the number of tokens. The results are presented in Table \ref{tab:perform-message-length}.

With short-length examples, the accuracy of most models (except Mistral) increased compared to the zero-shot learning results in Table \ref{tab:zeroshot} and the baseline few-shot learning in Table \ref{tab:perform-few-shot}. However, while most models did not surpass the performance of the few-shot learning results in Table \ref{tab:perform-message-number}, \textbf{Mixtral (8$\times$7B)} showed slight improvements, achieving an accuracy of 89.30\%, respectively.

Similarly, with medium-length examples, the accuracy of most models improved over the zero-shot learning results in Table \ref{tab:zeroshot}, except Mistral, which shows a decline in accuracy to 74.29\%. \textbf{Mixtral (8$\times$7B)} maintained consistent performance with only a slight increase in FNR. All models demonstrated significant improvements over the baseline few-shot models in Table \ref{tab:perform-few-shot}. While \textbf{LLAMA (70B)} and \textbf{Mixtral (8$\times$7B)} showed slight decreases in accuracy, all other models saw improvements compared to their performance in Table \ref{tab:perform-message-number}. Notably, \textbf{GPT-4} achieved its best performance in all experiments, with a precision of 97. 18\%, a nearly perfect TNR of 99.20\%, and a strong FNR of 4.97\%.

In contrast to the short and medium-length examples, the performance of the LLMs dropped significantly with lengthy messages. While \textbf{GPT-4} maintained a strong performance with an accuracy of 94.10\%, comparable to its few-shot learning results in Table \ref{tab:perform-message-number}, the performance of all other models decreased noticeably. Additionally, there were varying trends when compared to the zero-shot and baseline few-shot learning results. While \textbf{LLAMA (70B)}, \textbf{LLAMA (13B)}, and \textbf{GPT-4} showed slight improvements, the performance of other models declined significantly.

These findings suggest that while short and medium-length examples are effective for LLMs in few-shot learning, lengthy examples present a challenge, likely due to the increased complexity and variability of the content.

The ablation study highlights several key insights into the sensitivity of LLMs to the characteristics of few-shot learning examples in SMS spam detection. Our experiments reveal that the type and length of spam exemplars significantly influence the performance of LLMs in few-shot learning. Promotional examples generally yielded better results compared to fraud examples, particularly in terms of false positive rate (FPR). Short and medium-sized messages led to higher accuracies, with models like LLAMA (70B) and Mixtral (8$\times$7B) achieving near 90\% accuracy. DeepSeek (236B) also performed well with a high true negative rate (TNR), though it struggled with a lower true positive rate (TPR). Moreover, the consistent ratio of promotional to fraud messages within each length category confirmed that promotional messages, especially shorter ones, enhance LLMs' classification performance. In particular, LLAMA (70B), LLAMA (13B), and Mixtral (70B) achieved the best overall accuracy with short messages. In general, these experiments underscore the importance of carefully selecting and structuring few-shot learning examples to optimize the performance of LLMs in spam detection tasks. 

\subsection{Chain-of-thought LLM Prompting}
\label{sec:cot_eval}

Next, we explore the reasoning ability of various LLMs through chain-of-thought prompting. We explore whether the chain-of-thought prompting can further enhance few-shot learning with the optimal examples set. The Chain-of-thought prompt design can be seen in Figure \ref{fig:cot_prompt} in the appendix.

When employing chain-of-thought prompting, the models showed varied results in their performance metrics. LLAMA (70B) achieved an overall accuracy of 82.51\%, with its all time best (in this study) TPR of 92.48\% , indicating strong spam detection capabilities. However, its TNR was lower at 72.35\%, suggesting that while the model effectiveness improved at identifying spam, it struggled somewhat with accurately identifying legitimate messages. LLAMA (13B) showed a slightly lower accuracy of 78.69\%, with an impressive TPR of 97.39\%, but its TNR dropped to 59.21\%, further highlighting challenges in distinguishing non-spam content. Similarly, Mixtral (8$\times$7B) achieving the highest accuracy of 88.89\% among the open-source models in chain-of-thought, with its all-time best (in this study) TPR of 94.75\% and a TNR of 82.71\%. Mixtral (7B) had a lower overall accuracy of 71.27\%, with a TPR of 77.08\% and a TNR of 65.27\%, indicating that it was less effective compared to its larger counterpart. While the accuracy of LLAMA (70B), LLAMA (13B), and Mixtral (8$\times$7B) shows an improvement over their zero-shot performance in Table \ref{tab:zeroshot}, none of them exhibit further gains beyond the best accuracies attained during few-shot learning.

DeepSeek (236B) showed an overall performance with an accuracy of 93.45\%, coupled with a TPR of 90.39\% and a TNR of 96.50\%, indicating that it was highly effective at both detecting spam and correctly identifying legitimate messages. However, the standout performance came from GPT-4 (1.76T), which achieved the highest accuracy of 97.60\%, 
surpassing its previous high of 97.18\% (see Table \ref{tab:perform-message-length}). GPT-4 also maintained a balance with a TPR of 96.10\% and a TNR of 99.10\%, showcasing its ability to accurately classify both spam and non-spam content under adversarial conditions.

Our analysis indicates that certain models struggled to accurately adhere to the nuanced requirements set by the chain of thoughts prompt. For instance, Llama 7B and Mixtral 7B exhibited a tendency to misclassify benign messages as scams, suggesting a difficulty in differentiating casual language from potential scam indicators. This misalignment points towards a deficiency in these models' ability to contextualize informal language within the structured analysis framework. 

Conversely, DeepSeek and Mixtral 70B faced challenges in detecting scams that were cloaked in legitimate-looking texts, which often included subtle cues like embedded URLs or urgent calls to action. This highlights a shortfall in these models' capacity to scrutinize intent and verify source authenticity as outlined in the chain of thoughts.

The Llama 70B model, in particular, showed an over-sensitivity to emotionally charged language, often categorizing urgent but benign messages as scams (false positives). This over-sensitivity, while potentially beneficial in identifying certain types of scams, underscores a critical challenge in adhering to the directive to analyze the sender’s intent without bias, leading to significant misinterpretations.

When compared to a few-shot learning approach, which leverages a smaller set of examples to guide model predictions, the chain of thoughts approach generally resulted in improved performance across several metrics. Models like GPT-4 and DeepSeek showed marked improvements in both accuracy and specific metrics such as True Positive Rates (TPR) and True Negative Rates (TNR). 

GPT-4, in particular, demonstrated exceptional adaptability, benefiting from its extensive training on diverse datasets which likely enhanced its capability to process the detailed requirements of the chain of thoughts prompt. However, smaller models like LLAMA (13B) and Mistral (7B) exhibited inconsistencies, highlighting the limitations of few-shot learning when models lack extensive pre-training or face complex analytical tasks.

\subsection{Impact of Prompt Variability on LLM Performance Across Different Model Sizes}
\label{llm_size_impact}

\begin{figure}[th]
\centering
\includegraphics[width=1\linewidth]{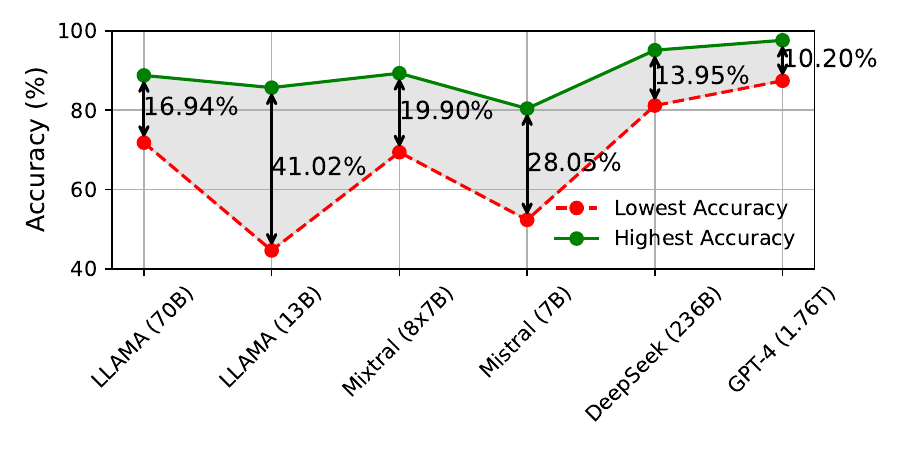}
\caption{Comparison of the highest and lowest accuracy of LLMS across prompts with highlighted differences.}
\label{fig:impact}
\end{figure}

The impact of prompts varies significantly across models, particularly between smaller and larger LLMs as shown in Figure \ref{fig:impact}. Prompt selection has the most significant effect on smaller LLMs. For the smaller LLAMA-2 (13B), the accuracy of the model with different prompts varies significantly, ranging from a maximum of 85.67\% (see Few-Shot Learning with Medium Size messages in Table \ref{tab:perform-message-length}) to a low of 44.65\% (see Few-Shot Learning with Fraud examples \ref{tab:perform-ml-fraud-promo}). This adjustment also resulted in a substantial decrease in its respective TNR from 75.61\% to 1.57\%, indicating that the model became less accurate in identifying negatives. However, the TPR remained comparatively high, above 95\%, showing that the model was still highly responsive to positive instances. Similarly, for the smaller Mixtral 7B model, adjusting the prompt in different settings led to a noticeable drop in accuracy, decreasing from a high of 80.40\% (see Table \ref{tab:zeroshot}) to 52.35\% (see Table \ref{tab:perform-few-shot}). This adjustment also resulted in a substantial decrease in their respective TNR from 63.44\% to 6.12\%, again indicating that the model became less accurate in identifying negatives. However, the TPR remained consistent at 100\%, showing that the model was still highly responsive to positive instances.

In contrast, larger models like GPT-4 (1.76T) show comparatively less variation in accuracy when exposed to different prompts. The accuracy for GPT-4 across varying conditions fluctuated between 87.40\% (see Table \ref{tab:perform-few-shot}) and 97.60\% (see Table \ref{tab:perform-chain}). This suggests that prompts have a lesser influence on the performance of larger models in comparison to smaller models.

Overall, while both smaller and larger models are sensitive to changes in prompts—evidenced by variation in accuracy, the impact on the smaller LLMs is the largest.


{\bf Takeaways.} The evaluation highlights that, while zero-shot learning is not reliable for SMS spam detection, few-shot learning approaches can be effective when data availability is limited. Incorporating few-shot examples significantly improved LLMs' performance. The type and length of the examples had a substantial impact on the results, with customized examples from specific categories and lengths providing better performance boosts. Moreover, in both zero-shot and all few-shot learning experiments, larger models consistently achieved higher accuracy compared to smaller models. Smaller models such as LLAMA (13B) and Mixtral (7B) exhibited greater sensitivity to the type, number, and length of a few-shot examples, and their performance varies significantly depending on the selected prompts. Chain-of-thought prompting generally improved reasoning abilities but did not uniformly improve performance across all models. In particular, GPT-4 in the commercial category and Mixtral (70B) in the open-source category outperformed the other LLMs. The findings underscore the importance of carefully tailoring few-shot examples and prompts, especially for smaller LLMs, to achieve optimal results in spam detection. The variability in performance across different models highlights that a one-size-fits-all approach is ineffective.

\section{Fine-tuning}
\label{sec:finetunning_eval}

Given that our zero-shot experiments did not produce \textit{satisfactory} performance for any LLM, while only GPT-4 achieved \textit{satisfactory} performance in few-shot learning, we decided to fine-tune the models to see if we could improve the results to \textit{Good} or \textit{Perfect} category. The other models did not produce good results, particularly in achieving balanced low FPR and FNR. Specifically, our goal was to reduce both FPR and FNR to below 3\% at minimum. 

To this end, we fine-tuned the models on the train split of our dataset and then evaluated their performance on the test split (refer to Section 5). The fine-tuning process was applied to the top three open-source LLMs, selected based on their highest achieved performance in zero-shot and few-shot learning. Specifically, LLAMA-2 (70B), LLAMA-2 (13B), and Mixtral (8$\times$7B) were chosen due to their higher accuracy levels among open-source LLMs, achieving 89.01\%, 85.57\%, and 89.30\%, respectively. The results of the fine-tuning experiments are presented in Table \ref{tab:finetuned}.

\begin{table}[hbt!]
\renewcommand{\arraystretch}{1.15}
\tabcolsep=0.05cm
\begin{center}
\caption{Performance evaluation of {\it Fine-tuned LLMs}. Fine-tuning significantly enhances LLM performance in SMS spam detection, with Mixtral achieving a balanced low FPR and FNR, meeting our criteria for a \textit{Good} spam detector.} 
\label{tab:finetuned}
\begin{tabular}{ c | c | c | c | c | c | c}
\toprule
 \hline
 & \multicolumn{6}{c}{\bf Performance Metrics}\\
\cline{2-7}
 \textbf{Model} & \textbf{Acc} & \textbf{FS} & \textbf{TPR} & \textbf{TNR} & \textbf{FPR} & \textbf{FNR} \\ [0.5ex]
 \midrule\hline
LLAMA-2 (70B) & 94.75\% & 94.45\% & 90.49\% & 99.96\% & 0.04\% & 9.51\% \\
LLAMA-2 (13B) & 92.10\% & 91.96\% & 88.00\% & 98.46\% & 1.54\% & 12.00\% \\
Mixtral (8$\times$7B) & 98.61\% & 98.57\% & 98.15\% & 99.04\% & 0.96\% & 1.85\% \\
  \hline
  \bottomrule
\end{tabular}
\end{center}
\end{table}

The fine-tuning results demonstrate a significant improvement in the performance of all the models compared to their zero-shot and few-shot learning counterparts. The Mixtral model, in particular, achieved an accuracy of 98.61\%, with both FPR and FNR below 2\%, meeting our desired criteria for a \textit{good} detector. Specifically, Mixtral (8$\times$7B) recorded an FPR of 0.96\% and an FNR of 1.85\%, indicating a well-balanced and robust performance in detecting spam messages. LLAMA-2 (70B) also showed marked improvement, achieving an accuracy of 94.75\%. While its FNR was 9.51\%, higher than our target, its FPR was exceptionally low at 0.04\%, demonstrating excellent precision in identifying non-spam messages. This disparity between TPR and TNR suggests that LLAMA-2 (70B) is particularly conservative in labeling messages as spam, potentially due to its tuning parameters. Similarly, LLAMA-2 (13B) demonstrated solid performance post-fine-tuning, with an accuracy of 92.10\%. Although its FNR was relatively higher at 12.00\%, its FPR remained low at 1.54\%, reflecting a balance towards cautious spam classification. While LLAMA-2 (70B) and LLAMA-2 (13B) demonstrated marked improvements in their performance, both failed to meet our minimum criteria for \textit{satisfactory} spam detection due to their high FNR (>5\%).

In general, the fine-tuning process significantly improved the performance of all models, with Mixtral (8$\times$7B) emerging as the superior model for SMS spam detection. These results underscore the importance of fine-tuning in leveraging the full potential of LLMs for specific tasks such as spam detection. Future work should continue to explore and optimize fine-tuning strategies to further enhance the robustness and accuracy of these models.

\section{LLMs Adversarial Resistance}
\label{sec:attacks_eval}

We investigate the adversarial robustness of LLMs in spam classification. An extensive set of experiments was conducted to profile the resistance of fine-tuned and pre-trained LLMs to various adversarial manipulations of SMS messages. The goal was to determine whether LLMs are robust to adversarial SMS spam attacks in general and to evaluate the impact of fine-tuning on their robustness.

Table \ref{tab:robust_llms} presents the results of extensive experiments comparing the performance of various LLMs in both zero-shot and fine-tuned settings across different types of adversarial attacks, using the original spam messages as a baseline. This analysis provides a comprehensive understanding of how different models perform under adversarial manipulations in SMS spam classification. The models are considered robust to adversarial manipulations if their accuracy under adversarial attacks does not fall below their baseline performance on the "Original" spam.

In the zero-shot setting, models were evaluated to assess their inherent robustness against adversarial attacks. LLAMA (70B) demonstrated strong performance, with accuracy ranging from 94.9\% to 98.5\%, indicating robust performance as it remained close to its original accuracy of 96.9\%. Similarly, LLAMA2 (13B) displayed robustness, with accuracy ranging from 96.4\% to 98.2\%, maintaining performance above its original accuracy of 94.9\%. The Mixtral (70B) also performed well, with accuracy ranging from 95.9\% to 99.5\% across different attack types, maintaining its robustness close to the original accuracy of 95.9\%. The Mistral (7B) model showed consistent performance, with accuracy ranging from 94.9\% to 97\%, indicating moderate robustness, close to its original accuracy of 94.9\%. Similarly, GPT-4 exhibited exceptional robustness, with accuracy rates consistently high between 98.7\% and 99.7\% across all attack types, maintaining performance above the original accuracy of 98.7\%. This highlights GPT-4's strong resilience to adversarial SMS spam attacks straight out of the box. In contrast, DeepSeek exhibited lower overall performance, with accuracy ranging from 81.6\% to 94.9\%, and in several attack categories, it fell below its original accuracy of 89.3\%, suggesting a vulnerability to adversarial attacks. DeepSeek's comparatively weaker performance against imperceptible adversarial manipulations especially \textit{Invisible} and {Reoerder} can be attributed to limitations in to its training data. DeepSeek, like other LLMs, relies heavily on its underlying architecture and the diversity of its training data. If DeepSeek's architecture or its pre-training data lacked sufficient exposure to adversarial patterns or linguistic variability, the model would be less equipped to handle the subtle perturbations introduced by adversarial manipulations. This shortfall could account for its vulnerability, particularly against imperceptible attacks such as non-printable characters. Without adequate training on these kinds of adversarial examples, DeepSeek may struggle to maintain accuracy when confronted with these nuanced attacks, leading to a drop in its overall performance.

In the case of fine-tuning, LLAMA (70B), which already performed well in the zero-shot setting, saw slight improvements, maintaining accuracy levels between 97.2\% and 98.9\%. Similarly, LLAMA (13B) showed consistent robustness after fine-tuning, with an accuracy ranging from 96.4\% to 98.2\%. Moreover, the fine-tuned Mixtral (8$\times$7B) model demonstrated significant enhancement, with accuracy rates ranging from 97.4\% to 100\%. The model achieved perfect accuracy in several categories, including spacing, insertion of characters, deletion of characters, swapping characters, and substitution of words. This confirms that fine-tuning can significantly bolster a model's resistance to adversarial manipulations.

\descr{Comparison with Traditional and Deep Machine Learning Models.} To provide context for the robustness of LLMs, it is essential to compare these results with conventional machine learning models. A recent study \cite{salman2024investigating} has shown that traditional ML models, such as Two-Class Classification (support vector machines (SVMs)), One-Class Classification (one-class SVM), Positive and Unlabeled (PU) Learning with random forests, neural networks, and state-of-the-art BERT-based architectures, are particularly susceptible to adversarial attacks. For example, in the study, all models experienced accuracy drops of over 40\% under the same adversarial manipulations, with \textit{spacing} and \textit{charswap} attacks being the most effective. Similarly, another recent study \cite{li2024spamdam} shows that BERT-based models experienced an accuracy drop of 14\% when exposed to imperceptible adversarial manipulations. This significant vulnerability is primarily due to the limited generalization capabilities of these models when confronted with perturbed input data. In contrast, LLMs, demonstrate markedly higher resilience, as evidenced by minimal accuracy drops under similar adversarial conditions.

These results highlight several key insights into the adversarial robustness of LLMs in SMS spam classification tasks. First, most LLM models exhibit strong inherent robustness in zero-shot settings, making them effective even without fine-tuning. However, this robustness is not universal across all models, as evidenced by the poorer performance of Deepseek against imperceptible attacks. Second, fine-tuning generally enhances the robustness of LLMs, particularly for models like Mixtral, where fine-tuning leads to perfect or near-perfect accuracy across many adversarial scenarios. This underscores the importance of fine-tuning in developing more resilient spam detection models. Finally, there is significant variability in how different LLMs respond to adversarial attacks, both in zero-shot and fine-tuned settings. This suggests that the choice of model and the application of fine-tuning are critical factors in achieving robust performance in adversarial environments.

\section{Concept Drift}
\label{sec:drift_eval}
Having evaluated the robustness of LLMs against adversarial attacks, we further investigate another critical robustness issue: concept drift. Concept drift refers to changes in the statistical properties of the target variable, which the model aims to predict, over time. In the context of SMS spam detection, this means that the nature of spam messages evolves, potentially rendering models trained on older data less effective on newer data.

While it is a well-known practice to address concept drift using periodic retraining of ML models, we aim to investigate how well LLMs can manage the concept drift and profile their performance in spam detection under these conditions. Specifically, we are interested in evaluating the impact of concept drift on LLM-based SMS spam detection to determine whether LLMs can maintain high detection rates without frequent retraining. To this end, we designed an experiment involving the fine-tuning of the Mixtral model on an outdated dataset and testing it on a more recent set of SMS messages. Specifically, we fine-tuned the Mixtral model using the widely used UCI SMS spam dataset \cite{almeida2011contributions}. After fine-tuning, we tested the model on a new set of SMS spam messages reported between 2020 and 2023.

We also conducted baseline experiments using traditional machine learning models with different architectures that have previously shown superior performance in SMS spam detection. These models include Support Vector Machine (SVM), Random Forest (RF), LightGBM (LGB), Long Short-Term Memory (LSTM), and Convolutional Neural Networks (CNN). The baseline models were trained on the same outdated dataset and tested on the latest set of SMS messages.

Table \ref{tab:baseline_concept_drift} presents the performance metrics of these baseline models in both pre-concept drift and post-concept drift scenarios. The pre-concept drift metrics show that these models performed exceptionally well on an 80:20 train-test split of the outdated dataset. However, their performance drastically reduced when exposed to the latest spam messages, highlighting the impact of concept drift. This indicates their lack of robustness to concept drift, which is a critical limitation in real-world spam detection scenarios. These results are consistent with a recent study conducted by Li et al. \cite{li2024spamdam}, where the accuracy of a BERT-based model dropped significantly from 99.03\% to 60.34\% when trained on the same UCI SMS spam dataset \cite{almeida2011contributions} and tested on a collection of the latest SMS spam.

\begin{table}[hbt!]
\renewcommand{\arraystretch}{1.15}
\tabcolsep=0.05cm
\begin{center}
\caption{Concept Drift Analysis of {\it Baseline Models}. The baseline models show significant performance degradation due to concept drift, highlighting their lack of robustness in real-world spam detection scenarios.}
\label{tab:baseline_concept_drift}
\scalebox{0.80}{%
\begin{tabular}{ c | c | c | c | c | c | c | c | c | c | c }
\toprule
 \hline
 & \multicolumn{5}{c|}{\bf Pre-concept Drift} & \multicolumn{5}{c}{\bf Post-concept Drift} \\
\cline{2-11}
 \textbf{Model} & \textbf{Acc} & \textbf{TPR} & \textbf{TNR} & \textbf{FPR} & \textbf{FNR} & \textbf{Acc} & \textbf{TPR} & \textbf{TNR} & \textbf{FPR} & \textbf{FNR} \\ [0.5ex]
 \hline\hline
SVM & 99.73\% & 97.99\% & 100.00\% & 0.00\% & 2.01\% & 53.45\% & 7.20\% & 99.70\% & 0.30\% & 92.80\% \\
RF & 99.98\% & 99.87\% & 100.00\% & 0.00\% & 0.13\% & 57.50\% & 15.20\% & 99.80\% & 0.20\% & 84.80\% \\
LGB & 99.26\% & 94.64\% & 99.98\% & 0.02\% & 5.36\% & 62.50\% & 26.60\% & 98.40\% & 1.60\% & 73.40\% \\
LSTM & 99.26\% & 94.64\% & 99.98\% & 0.02\% & 5.36\% & 64.20\% & 29.20\% & 99.20\% & 0.80\% & 70.80\% \\
CNN & 99.26\% & 94.64\% & 99.98\% & 0.02\% & 5.36\% & 64.20\% & 29.20\% & 99.20\% & 0.80\% & 70.80\% \\
  \hline
  \bottomrule
\end{tabular}}
\end{center}
\end{table}

Table \ref{tab:mixtral_concept_drift} presents the performance metrics of the Mixtral (70B) model under these conditions. Despite being fine-tuned on older UCI SMS spam dataset \cite{almeida2011contributions}, the Mixtral model achieved an accuracy of 90.40\%, with a high true negative rate (TNR) of 99.70\%, indicating its ability to correctly identify legitimate messages. However, the true positive rate (TPR) was 81.10\%, suggesting some difficulties in detecting all spam messages, likely due to changes in spam characteristics over time. The false positive rate (FPR) remained low at 0.30\%, but the false negative rate (FNR) was 18.90\%, highlighting the impact of concept drift on spam detection accuracy. The Mixtral LLM, despite some performance drop, maintained relatively high accuracy and low false positive rates even with concept drift. This highlights the potential of LLMs to be more resilient to evolving spam characteristics compared to traditional models.

\begin{table}[hbt!]
\renewcommand{\arraystretch}{1.15}
\tabcolsep=0.05cm
\begin{center}
\caption{Concept Drift Analysis of {\it Mixtral (8$\times$7B)}.} 
\label{tab:mixtral_concept_drift}
\begin{tabular}{ c | c | c | c | c | c }
\toprule
 & \multicolumn{5}{c}{\bf Performance Metrics}\\
\cline{2-6}
 \textbf{Model} & \textbf{Acc} & \textbf{TPR} & \textbf{TNR} & \textbf{FPR} & \textbf{FNR} \\ [0.5ex]
 \hline\hline
Mixtral (Finetuned zero-shot) & 90.4\%  & 81.1\% & 99.7\% & 0.3\% & 18.9\% \\
Mixtral (Finetuned few-shot) & 96.2\% & 92.8\% & 99.5\% & 0.5\% & 7.2\% \\
  \bottomrule
\end{tabular}
\end{center}
\end{table}

To further investigate the model's ability to mitigate concept drift, we applied few-shot learning to the fine-tuned model by selecting five representative spam examples from the most recent spam data. The results, presented in Table \ref{tab:mixtral_concept_drift}, demonstrate that this strategy significantly enhanced the model's capacity to manage concept drift. The Mixtral (Finetuned few-shot) model achieved an accuracy of 96.15\%, with a true positive rate (TPR) of 92.80\%, a true negative rate (TNR) of 99.50\%, and a reduced false negative rate (FNR) of 7.20\% although a little higher than the \textit{satisfactory} threshold of 5\%. These improvements suggest that incorporating few-shot learning into the fine-tuning process enables the model to better adapt to the evolving characteristics of spam, thereby maintaining high detection rates even as spam tactics change over time. This approach highlights the potential of few-shot fine-tuning as an effective strategy for combating concept drift in SMS spam detection, especially in scenarios where retraining or extensive fine-tuning on large, updated datasets is impractical due to resource constraints or data availability.

These findings underscore the critical importance of regularly updating and retraining spam detection models to maintain high performance in the face of evolving spam tactics. The significant improvement in performance achieved by the Mixtral (Finetuned few-shot) model highlights the potential of few-shot learning as an effective approach to mitigating the impact of concept drift, particularly when extensive retraining on large datasets is not feasible. While traditional models showed a drastic drop in true positive rates (TPR) when tested on newer data, the few-shot fine-tuning approach enabled the Mixtral model to adapt more effectively to recent spam characteristics, maintaining higher detection rates. Future work should explore the development of continuous learning and adaptive strategies, such as automated retraining pipelines and few-shot fine-tuning techniques, to address concept drift in real-time and ensure sustained model robustness against the ever-changing landscape of spam behaviors.

\section{Discussion}
We comprehensively evaluated the effectiveness of LMs for addressing the challenges in SMS spam detection, focusing on zero-shot, few-shot, chain-of-thought, and fine-tuning learning approaches. Our findings reveal several critical insights into the capabilities and limitations of these models, providing a foundation for future improvements and applications in spam detection.

Our zero-shot learning experiments highlighted the inherent challenges in using LLMs without task-specific training. Although larger models such as Mixtral-70B and commercial models such as DeepSeek (236B) and GPT-4 outperformed their smaller counterparts, none of them achieve \textit{satisfactory} spam detection. This indicates that despite the extensive pre-training on diverse datasets, LLMs require further adaptation to effectively distinguish between spam and legitimate messages in a zero-shot context.

Few-shot learning significantly improved the performance of most LLMs. Our experiments demonstrated that carefully selected examples from various spam categories can enhance model performance; however, challenges remain in balancing TPR and TNR. Achieving this balance is crucial to minimize both FPR and FNR, which is essential for reliable spam detection. Despite comprehensive tuning with different examples, only GPT-4 achieved \textit{satisfactory} spam detection. These findings suggest that while careful selection of examples is essential for optimizing model performance, few-shot learning alone may not be sufficient to achieve the best possible outcomes in spam detection. Additionally, our analysis of few-shot learning reveals insights into the transferability of system prompts across different LLMs and their varying impact on small and large models. We found that a system prompt that works well on one LLM does not guarantee optimal performance on another, indicating poor transferability of developed prompts across different models. Moreover, our analysis indicates that smaller LLMs are more sensitive to system prompts compared to larger LLMs, exhibiting greater variations in performance when exposed to different prompts.

Fine-tuning emerged as the most effective method for optimizing LLMs for SMS spam detection. Fine-tuning significantly enhances the performance of all three LLMs (Section \ref{sec:finetunning_eval}), particularly Mixtral-70B, which achieved an impressive accuracy of 98.7\%, with both FPR and FNR below 2\% getting into the \textit{Good} category of spam detection. This underscores the importance of adapting pre-trained models to specific tasks through fine-tuning. The fine-tuning process allows the models to learn the nuances of the characteristics of spam, significantly improving their detection capabilities. 

The robustness of LLMs against adversarial manipulations was another focal point of our study. We evaluated the models against various adversarial attacks, including conventional perceptible evasion manipulations like character perturbations and sophisticated, imperceptible attacks involving non-printable Unicode characters. Our results showed that while most LLMs demonstrated exceptional robustness against these manipulations, both in their pre-trained (zero-shot) and fine-tuned states, this robustness is not universal across all models. The variability in performance across different LLMs and settings underscores the need for careful consideration when selecting and fine-tuning models for adversarially resistant spam detection systems. This robustness is essential for maintaining high accuracy in real-world applications, where adversaries continuously evolve their tactics to bypass spam filters.

We also investigated the impact of concept drift, where the nature of spam messages evolves over time, potentially reducing the effectiveness of static models. Our experiments with the Mixtral-70B model, which demonstrated the highest accuracy among the fine-tuned models, revealed that LLMs are more resilient to concept drift compared to traditional machine-learning models. The model's accuracy further improved when the latest examples of spam were incorporated into the few-shot learning process. This resilience suggests that LLMs could be a more reliable choice for long-term spam detection, reducing the frequency of retraining required to maintain high performance.

In summary, our analysis underscores the significant potential of LLMs in SMS spam detection, particularly when models are fine-tuned and rigorously evaluated against adversarial threats and concept drift. Although zero-shot and few-shot learning provides valuable insights into model capabilities, fine-tuning is essential for achieving optimal performance. The substantial performance gains observed through fine-tuning highlight its necessity for the effective deployment of LLMs in real-world spam detection applications.  

\section{Related Work}
To address the threat posed by SMS spam, several ML learning based techniques have been proposed. 
These methods encompass traditional classification techniques such as support vector machines (SVM)\cite{almeida2011contributions}, and random forest \cite{sjarif2019sms}, deep learning architectures such as convolutional neural networks (CNNs)\cite{ghourabi2020hybrid}, long short-term memory networks (LSTMs)\cite{ghourabi2020hybrid}, and bidirectional LSTM (BiLSTM)\cite{liu2021spam} and transformer models \cite{liu2021spam} as well as the latest transformer architectures such as Bidirectional Encoder Representations from Transformers (BERT) \cite{oswald2022spotspam,sahmoud2022spam}.
However, all of these methods are prone to adversarial perturbations and concept drift. 

Salman et al. \cite{salman2024investigating} evaluated a large number of machine learning models, including conventional two-class machine learning models such as SVM, one-class ML models, PU learning and BERT-based architectures, as well as ML-based mobile text spam filtering apps and anti-spam services. They found that all of these models are vulnerable to several black-box manipulations such as paraphrasing, insertion, replacement, deletion, and homographs. Additionally, Salman et al. \cite{salman2024investigating} performed a concept drift analysis of the ML models and found a significant drop in the accuracies of the models when trained on an older dataset and tested against new SMS spam. Similarly, Li et al. \cite{li2024spamdam} evaluated the robustness of a BERT-trained SMS spam detection model against a new set of imperceptible adversarial attacks (Boucher et al. \cite{boucher2022bad}) and found that all attacks successfully evaded the model. Boucher et al. \cite{boucher2022bad} introduced a method for generating imperceptible adversarial attacks by either adding non-printable Unicode characters or substituting existing characters with visually similar homoglyphs, making the adversarial modifications undetectable. We adopted these imperceptible adversarial attacks along with the conventional perceptible black-box attacks evaluated by Salman et al. \cite{salman2024investigating}.

\section{Conclusion}
We performed an extensive evaluation of various LLMs for SMS spam detection, examining their performance under zero-shot, few-shot, chain-of-thought, and fine-tuning scenarios. Our findings reveal that while zero-shot learning offers convenience, it is generally unreliable for effective spam detection. Few-shot learning shows promise, particularly when examples are carefully selected based on the model’s sensitivities, but exhibits significant variability depending on the model and the characteristics of the examples used. Despite the improvement in performance by most LLMs, only the commercial GPT-4 meets our ``satisfactory'' criteria for spam detection, and none of the LLMs achieve the ``Good'' or ``Perfect'' category. 

The chain-of-thought prompting approach slightly enhances performance in some cases, although not uniformly across all models, leaving room for improvement.

Moreover, our analysis of adversarial robustness reveals that LLMs inherently possess a certain degree of resistance to adversarial attacks. Fine-tuning further enhances this robustness, as demonstrated by the Mixtral (8$\times$7B) model, which achieved near-perfect accuracy against various adversarial manipulations post-fine-tuning. This suggests that LLMs can be highly resilient to adversarial attempts to manipulate spam detection systems, a challenge that traditional spam detectors still face. 

Fine-tuning the models shows significant improvement for all three open-source LLMs, with Mixtral (8$\times$7B) particularly achieving a balanced FPR (<1\%) and FNR (<2\%), satisfying the criteria for ``Good'' spam detection. 
Importantly, our experiments demonstrate that fine-tuned LLMs, particularly Mixtral (8$\times$7B), maintain high performance even when tested on datasets significantly different (newer data) from those on which they were trained (older data). This indicates that LLMs are capable of mitigating the impact of concept drift more effectively than traditional models, potentially reducing the need for frequent retraining. 

Overall, our study demonstrates that while LLMs hold significant potential for SMS spam detection, particularly with fine-tuning, their deployment in real-world scenarios requires careful consideration of model selection, training strategies, and data characteristics. The ability of LLMs to address adversarial robustness and mitigate concept drift positions them as promising tools to maintain effective spam detection in dynamic and adversarial environments in the real world. 

Future work should focus on optimizing fine-tuning techniques and further exploring advanced prompting methods to enhance the robustness and generalizability of LLMs across varied and evolving spam landscapes.

%
%
%
\bibliographystyle{ACM-Reference-Format}
\bibliography{main}


\begin{thebibliography}{40}


\ifx \showCODEN    \undefined \def \showCODEN     #1{\unskip}     \fi
\ifx \showISBNx    \undefined \def \showISBNx     #1{\unskip}     \fi
\ifx \showISBNxiii \undefined \def \showISBNxiii  #1{\unskip}     \fi
\ifx \showISSN     \undefined \def \showISSN      #1{\unskip}     \fi
\ifx \showLCCN     \undefined \def \showLCCN      #1{\unskip}     \fi
\ifx \shownote     \undefined \def \shownote      #1{#1}          \fi
\ifx \showarticletitle \undefined \def \showarticletitle #1{#1}   \fi
\ifx \showURL      \undefined \def \showURL       {\relax}        \fi
\providecommand\bibfield[2]{#2}
\providecommand\bibinfo[2]{#2}
\providecommand\natexlab[1]{#1}
\providecommand\showeprint[2][]{arXiv:#2}

\bibitem[act(2023)]%
        {actionfraud}
 \bibinfo{year}{2023}\natexlab{}.
\newblock \bibinfo{title}{Action Fraud}.
\newblock \bibinfo{howpublished}{\url{https://www.actionfraud.police.uk/}}.
\newblock
\newblock
\shownote{Last accessed 06 Oct 2023}.


\bibitem[ACCS(2022)]%
        {accs}
\bibfield{author}{\bibinfo{person}{ACCS}.} \bibinfo{year}{2022}\natexlab{}.
\newblock \bibinfo{title}{ACCS Scam statistics}.
\newblock \bibinfo{howpublished}{\url{https://www.scamwatch.gov.au/scam-statistics}}.
\newblock
\newblock
\shownote{Last accessed 15 Jul 2024]}.


\bibitem[Achiam et~al\mbox{.}(2023)]%
        {achiam2023gpt}
\bibfield{author}{\bibinfo{person}{Josh Achiam}, \bibinfo{person}{Steven Adler}, \bibinfo{person}{Sandhini Agarwal}, \bibinfo{person}{Lama Ahmad}, \bibinfo{person}{Ilge Akkaya}, \bibinfo{person}{Florencia~Leoni Aleman}, \bibinfo{person}{Diogo Almeida}, \bibinfo{person}{Janko Altenschmidt}, \bibinfo{person}{Sam Altman}, \bibinfo{person}{Shyamal Anadkat}, {et~al\mbox{.}}} \bibinfo{year}{2023}\natexlab{}.
\newblock \showarticletitle{Gpt-4 technical report}.
\newblock \bibinfo{journal}{\emph{arXiv preprint arXiv:2303.08774}} (\bibinfo{year}{2023}).
\newblock


\bibitem[Almeida et~al\mbox{.}(2013)]%
        {almeida2013towards}
\bibfield{author}{\bibinfo{person}{Tiago Almeida} {et~al\mbox{.}}} \bibinfo{year}{2013}\natexlab{}.
\newblock \showarticletitle{Towards sms spam filtering: Results under a new dataset}.
\newblock \bibinfo{journal}{\emph{{JiSS}}} \bibinfo{volume}{2}, \bibinfo{number}{1} (\bibinfo{year}{2013}).
\newblock


\bibitem[Almeida et~al\mbox{.}(2011)]%
        {almeida2011contributions}
\bibfield{author}{\bibinfo{person}{Tiago~A Almeida}, \bibinfo{person}{Jos{\'e} Mar{\'\i}a~G Hidalgo}, {and} \bibinfo{person}{Akebo Yamakami}.} \bibinfo{year}{2011}\natexlab{}.
\newblock \showarticletitle{Contributions to the study of SMS spam filtering: new collection and results}. In \bibinfo{booktitle}{\emph{Proceedings of the 11th ACM symposium on Document engineering}}. \bibinfo{pages}{259--262}.
\newblock


\bibitem[Boucher et~al\mbox{.}(2022)]%
        {boucher2022bad}
\bibfield{author}{\bibinfo{person}{Nicholas Boucher}, \bibinfo{person}{Ilia Shumailov}, \bibinfo{person}{Ross Anderson}, {and} \bibinfo{person}{Nicolas Papernot}.} \bibinfo{year}{2022}\natexlab{}.
\newblock \showarticletitle{Bad characters: Imperceptible nlp attacks}. In \bibinfo{booktitle}{\emph{2022 IEEE Symposium on Security and Privacy (SP)}}. IEEE, \bibinfo{pages}{1987--2004}.
\newblock


\bibitem[Brown et~al\mbox{.}(2020)]%
        {brown2020language}
\bibfield{author}{\bibinfo{person}{Tom Brown}, \bibinfo{person}{Benjamin Mann}, \bibinfo{person}{Nick Ryder}, \bibinfo{person}{Melanie Subbiah}, \bibinfo{person}{Jared~D Kaplan}, \bibinfo{person}{Prafulla Dhariwal}, \bibinfo{person}{Arvind Neelakantan}, \bibinfo{person}{Pranav Shyam}, \bibinfo{person}{Girish Sastry}, \bibinfo{person}{Amanda Askell}, {et~al\mbox{.}}} \bibinfo{year}{2020}\natexlab{}.
\newblock \showarticletitle{Language models are few-shot learners}.
\newblock \bibinfo{journal}{\emph{Advances in neural information processing systems}}  \bibinfo{volume}{33} (\bibinfo{year}{2020}), \bibinfo{pages}{1877--1901}.
\newblock


\bibitem[Chen and Kan(2013)]%
        {chen2013creating}
\bibfield{author}{\bibinfo{person}{Tao Chen} {and} \bibinfo{person}{Min-Yen Kan}.} \bibinfo{year}{2013}\natexlab{}.
\newblock \showarticletitle{Creating a live, public short message service corpus: the NUS SMS corpus}.
\newblock \bibinfo{journal}{\emph{LRE}} (\bibinfo{year}{2013}).
\newblock


\bibitem[{DeepSeek}(2023)]%
        {deepseek}
\bibfield{author}{\bibinfo{person}{{DeepSeek}}.} \bibinfo{year}{2023}\natexlab{}.
\newblock \bibinfo{title}{DeepSeek API}.
\newblock \bibinfo{howpublished}{\url{https://platform.deepseek.com/api-docs/api/deepseek-api/}}.
\newblock
\newblock
\shownote{Last accessed 15 Jul 2024]}.


\bibitem[Dettmers et~al\mbox{.}(2024)]%
        {dettmers2024qlora}
\bibfield{author}{\bibinfo{person}{Tim Dettmers}, \bibinfo{person}{Artidoro Pagnoni}, \bibinfo{person}{Ari Holtzman}, {and} \bibinfo{person}{Luke Zettlemoyer}.} \bibinfo{year}{2024}\natexlab{}.
\newblock \showarticletitle{Qlora: Efficient finetuning of quantized llms}.
\newblock \bibinfo{journal}{\emph{Advances in Neural Information Processing Systems}}  \bibinfo{volume}{36} (\bibinfo{year}{2024}).
\newblock


\bibitem[Dunlap et~al\mbox{.}(2024)]%
        {dunlap2024pairing}
\bibfield{author}{\bibinfo{person}{Trevor Dunlap}, \bibinfo{person}{John~Speed Meyers}, \bibinfo{person}{Bradley Reaves}, {and} \bibinfo{person}{William Enck}.} \bibinfo{year}{2024}\natexlab{}.
\newblock \showarticletitle{Pairing Security Advisories with Vulnerable Functions Using Open-Source LLMs}. In \bibinfo{booktitle}{\emph{International Conference on Detection of Intrusions and Malware, and Vulnerability Assessment}}. Springer, \bibinfo{pages}{350--369}.
\newblock


\bibitem[Fatemi and Hu(2023)]%
        {fatemi2023comparative}
\bibfield{author}{\bibinfo{person}{Sorouralsadat Fatemi} {and} \bibinfo{person}{Yuheng Hu}.} \bibinfo{year}{2023}\natexlab{}.
\newblock \showarticletitle{A Comparative Analysis of Fine-Tuned LLMs and Few-Shot Learning of LLMs for Financial Sentiment Analysis}.
\newblock \bibinfo{journal}{\emph{arXiv preprint arXiv:2312.08725}} (\bibinfo{year}{2023}).
\newblock


\bibitem[Fields et~al\mbox{.}(2024)]%
        {fields2024survey}
\bibfield{author}{\bibinfo{person}{John Fields}, \bibinfo{person}{Kevin Chovanec}, {and} \bibinfo{person}{Praveen Madiraju}.} \bibinfo{year}{2024}\natexlab{}.
\newblock \showarticletitle{A survey of text classification with transformers: How wide? how large? how long? how accurate? how expensive? how safe?}
\newblock \bibinfo{journal}{\emph{IEEE Access}} (\bibinfo{year}{2024}).
\newblock


\bibitem[Ge et~al\mbox{.}(2024)]%
        {ge2024openagi}
\bibfield{author}{\bibinfo{person}{Yingqiang Ge}, \bibinfo{person}{Wenyue Hua}, \bibinfo{person}{Kai Mei}, \bibinfo{person}{Juntao Tan}, \bibinfo{person}{Shuyuan Xu}, \bibinfo{person}{Zelong Li}, \bibinfo{person}{Yongfeng Zhang}, {et~al\mbox{.}}} \bibinfo{year}{2024}\natexlab{}.
\newblock \showarticletitle{Openagi: When llm meets domain experts}.
\newblock \bibinfo{journal}{\emph{Advances in Neural Information Processing Systems}}  \bibinfo{volume}{36} (\bibinfo{year}{2024}).
\newblock


\bibitem[Ghourabi et~al\mbox{.}(2020)]%
        {ghourabi2020hybrid}
\bibfield{author}{\bibinfo{person}{Abdallah Ghourabi}, \bibinfo{person}{Mahmood~A Mahmood}, {and} \bibinfo{person}{Qusay~M Alzubi}.} \bibinfo{year}{2020}\natexlab{}.
\newblock \showarticletitle{A hybrid CNN-LSTM model for SMS spam detection in arabic and english messages}.
\newblock \bibinfo{journal}{\emph{Future Internet}} \bibinfo{volume}{12}, \bibinfo{number}{9} (\bibinfo{year}{2020}), \bibinfo{pages}{156}.
\newblock


\bibitem[Heimerl et~al\mbox{.}(2014)]%
        {heimerl2014word}
\bibfield{author}{\bibinfo{person}{Florian Heimerl} {et~al\mbox{.}}} \bibinfo{year}{2014}\natexlab{}.
\newblock \showarticletitle{Word cloud explorer: Text analytics based on word clouds}. In \bibinfo{booktitle}{\emph{{ICSS}}}.
\newblock


\bibitem[Hu et~al\mbox{.}(2021)]%
        {hu2021lora}
\bibfield{author}{\bibinfo{person}{Edward~J Hu}, \bibinfo{person}{Yelong Shen}, \bibinfo{person}{Phillip Wallis}, \bibinfo{person}{Zeyuan Allen-Zhu}, \bibinfo{person}{Yuanzhi Li}, \bibinfo{person}{Shean Wang}, \bibinfo{person}{Lu Wang}, {and} \bibinfo{person}{Weizhu Chen}.} \bibinfo{year}{2021}\natexlab{}.
\newblock \showarticletitle{Lora: Low-rank adaptation of large language models}.
\newblock \bibinfo{journal}{\emph{arXiv preprint arXiv:2106.09685}} (\bibinfo{year}{2021}).
\newblock


\bibitem[Jiang et~al\mbox{.}(2024)]%
        {jiang2024mixtral}
\bibfield{author}{\bibinfo{person}{Albert~Q Jiang}, \bibinfo{person}{Alexandre Sablayrolles}, \bibinfo{person}{Antoine Roux}, \bibinfo{person}{Arthur Mensch}, \bibinfo{person}{Blanche Savary}, \bibinfo{person}{Chris Bamford}, \bibinfo{person}{Devendra~Singh Chaplot}, \bibinfo{person}{Diego de~las Casas}, \bibinfo{person}{Emma~Bou Hanna}, \bibinfo{person}{Florian Bressand}, {et~al\mbox{.}}} \bibinfo{year}{2024}\natexlab{}.
\newblock \showarticletitle{Mixtral of experts}.
\newblock \bibinfo{journal}{\emph{arXiv preprint arXiv:2401.04088}} (\bibinfo{year}{2024}).
\newblock


\bibitem[Kalyan(2023)]%
        {kalyan2023survey}
\bibfield{author}{\bibinfo{person}{Katikapalli~Subramanyam Kalyan}.} \bibinfo{year}{2023}\natexlab{}.
\newblock \showarticletitle{A survey of GPT-3 family large language models including ChatGPT and GPT-4}.
\newblock \bibinfo{journal}{\emph{Natural Language Processing Journal}} (\bibinfo{year}{2023}), \bibinfo{pages}{100048}.
\newblock


\bibitem[Li et~al\mbox{.}(2024)]%
        {li2024spamdam}
\bibfield{author}{\bibinfo{person}{Yekai Li}, \bibinfo{person}{Rufan Zhang}, \bibinfo{person}{Wenxin Rong}, {and} \bibinfo{person}{Xianghang Mi}.} \bibinfo{year}{2024}\natexlab{}.
\newblock \showarticletitle{SpamDam: Towards Privacy-Preserving and Adversary-Resistant SMS Spam Detection}.
\newblock \bibinfo{journal}{\emph{arXiv preprint arXiv:2404.09481}} (\bibinfo{year}{2024}).
\newblock


\bibitem[Liu et~al\mbox{.}(2022)]%
        {liu2022few}
\bibfield{author}{\bibinfo{person}{Haokun Liu}, \bibinfo{person}{Derek Tam}, \bibinfo{person}{Mohammed Muqeeth}, \bibinfo{person}{Jay Mohta}, \bibinfo{person}{Tenghao Huang}, \bibinfo{person}{Mohit Bansal}, {and} \bibinfo{person}{Colin~A Raffel}.} \bibinfo{year}{2022}\natexlab{}.
\newblock \showarticletitle{Few-shot parameter-efficient fine-tuning is better and cheaper than in-context learning}.
\newblock \bibinfo{journal}{\emph{Advances in Neural Information Processing Systems}}  \bibinfo{volume}{35} (\bibinfo{year}{2022}), \bibinfo{pages}{1950--1965}.
\newblock


\bibitem[Liu et~al\mbox{.}(2021)]%
        {liu2021spam}
\bibfield{author}{\bibinfo{person}{Xiaoxu Liu}, \bibinfo{person}{Haoye Lu}, {and} \bibinfo{person}{Amiya Nayak}.} \bibinfo{year}{2021}\natexlab{}.
\newblock \showarticletitle{A spam transformer model for SMS spam detection}.
\newblock \bibinfo{journal}{\emph{IEEE Access}}  \bibinfo{volume}{9} (\bibinfo{year}{2021}), \bibinfo{pages}{80253--80263}.
\newblock


\bibitem[{Meta Llama}(2023a)]%
        {llama13B}
\bibfield{author}{\bibinfo{person}{{Meta Llama}}.} \bibinfo{year}{2023}\natexlab{a}.
\newblock \bibinfo{title}{Llama 2 13B}.
\newblock \bibinfo{howpublished}{\url{https://huggingface.co/meta-llama/Llama-2-13b-chat-hf}}.
\newblock
\newblock
\shownote{Last accessed 15 Jul 2024]}.


\bibitem[{Meta Llama}(2023b)]%
        {llama70B}
\bibfield{author}{\bibinfo{person}{{Meta Llama}}.} \bibinfo{year}{2023}\natexlab{b}.
\newblock \bibinfo{title}{Llama 2 70B}.
\newblock \bibinfo{howpublished}{\url{https://huggingface.co/meta-llama/Llama-2-70b-chat-hf}}.
\newblock
\newblock
\shownote{Last accessed 15 Jul 2024]}.


\bibitem[{Mistral AI}(2023a)]%
        {mistral}
\bibfield{author}{\bibinfo{person}{{Mistral AI}}.} \bibinfo{year}{2023}\natexlab{a}.
\newblock \bibinfo{title}{Model Card for Mistral7B}.
\newblock \bibinfo{howpublished}{\url{https://huggingface.co/mistralai/Mistral-7B-Instruct-v0.3}}.
\newblock
\newblock
\shownote{Last accessed 15 Jul 2024]}.


\bibitem[{Mistral AI}(2023b)]%
        {mixtral}
\bibfield{author}{\bibinfo{person}{{Mistral AI}}.} \bibinfo{year}{2023}\natexlab{b}.
\newblock \bibinfo{title}{Model Card for Mixtral-8x7B}.
\newblock \bibinfo{howpublished}{\url{https://huggingface.co/mistralai/Mixtral-8x7B-Instruct-v0.1}}.
\newblock
\newblock
\shownote{Last accessed 15 Jul 2024]}.


\bibitem[Morris et~al\mbox{.}(2020)]%
        {morris2020textattack}
\bibfield{author}{\bibinfo{person}{John~X Morris}, \bibinfo{person}{Eli Lifland}, \bibinfo{person}{Jin~Yong Yoo}, \bibinfo{person}{Jake Grigsby}, \bibinfo{person}{Di Jin}, {and} \bibinfo{person}{Yanjun Qi}.} \bibinfo{year}{2020}\natexlab{}.
\newblock \showarticletitle{Textattack: A framework for adversarial attacks, data augmentation, and adversarial training in nlp}.
\newblock \bibinfo{journal}{\emph{arXiv preprint arXiv:2005.05909}} (\bibinfo{year}{2020}).
\newblock


\bibitem[nickboucher(2022)]%
        {imperceptible_perturb}
\bibfield{author}{\bibinfo{person}{nickboucher}.} \bibinfo{year}{2022}\natexlab{}.
\newblock \bibinfo{title}{Imperceptible Perturbations}.
\newblock \bibinfo{howpublished}{\url{https://github.com/nickboucher/imperceptible}}.
\newblock
\newblock
\shownote{Last accessed 15 Jul 2024]}.


\bibitem[{OpenAI}(2023)]%
        {gpt4}
\bibfield{author}{\bibinfo{person}{{OpenAI}}.} \bibinfo{year}{2023}\natexlab{}.
\newblock \bibinfo{title}{Models - OpenAI API}.
\newblock \bibinfo{howpublished}{\url{https://platform.openai.com/docs/models/gpt-4-turbo-and-gpt-4}}.
\newblock
\newblock
\shownote{Last accessed 15 Jul 2024]}.


\bibitem[OpenAI(2024a)]%
        {prompteng}
\bibfield{author}{\bibinfo{person}{OpenAI}.} \bibinfo{year}{2024}\natexlab{a}.
\newblock \bibinfo{title}{Best practices for prompt engineering with the OpenAI API}.
\newblock \bibinfo{howpublished}{\url{https://help.openai.com/en/articles/6654000-best-practices-for-prompt-engineering-with-the-openai-api}}.
\newblock
\newblock
\shownote{Last accessed 16 Jul 2024]}.


\bibitem[OpenAI(2024b)]%
        {openAiApple}
\bibfield{author}{\bibinfo{person}{OpenAI}.} \bibinfo{year}{2024}\natexlab{b}.
\newblock \bibinfo{title}{OpenAI and Apple Announce Partnership}.
\newblock \bibinfo{howpublished}{\url{https://openai.com/index/openai-and-apple-announce-partnership/}}.
\newblock
\newblock
\shownote{Last accessed 15 Jul 2024]}.


\bibitem[Oswald et~al\mbox{.}(2022)]%
        {oswald2022spotspam}
\bibfield{author}{\bibinfo{person}{C Oswald}, \bibinfo{person}{Sona~Elza Simon}, {and} \bibinfo{person}{Arnab Bhattacharya}.} \bibinfo{year}{2022}\natexlab{}.
\newblock \showarticletitle{Spotspam: Intention analysis--driven sms spam detection using bert embeddings}.
\newblock \bibinfo{journal}{\emph{ACM Transactions on the Web (TWEB)}} \bibinfo{volume}{16}, \bibinfo{number}{3} (\bibinfo{year}{2022}), \bibinfo{pages}{1--27}.
\newblock


\bibitem[{Robokiller}(2023)]%
        {robo23}
\bibfield{author}{\bibinfo{person}{{Robokiller}}.} \bibinfo{year}{2023}\natexlab{}.
\newblock \bibinfo{title}{2023 Mid-Year Phone Scam Report}.
\newblock \bibinfo{howpublished}{\url{https://www.robokiller.com/robokiller-2023-mid-year-phone-scam-report}}.
\newblock
\newblock
\shownote{Last accessed 15 Jul 2024]}.


\bibitem[Sahmoud and Mikki(2022)]%
        {sahmoud2022spam}
\bibfield{author}{\bibinfo{person}{Thaer Sahmoud} {and} \bibinfo{person}{Dr~Mohammad Mikki}.} \bibinfo{year}{2022}\natexlab{}.
\newblock \showarticletitle{Spam detection using BERT}.
\newblock \bibinfo{journal}{\emph{arXiv preprint arXiv:2206.02443}} (\bibinfo{year}{2022}).
\newblock


\bibitem[Salman et~al\mbox{.}(2024)]%
        {salman2024investigating}
\bibfield{author}{\bibinfo{person}{Muhammad Salman}, \bibinfo{person}{Muhammad Ikram}, {and} \bibinfo{person}{Mohamed~Ali Kaafar}.} \bibinfo{year}{2024}\natexlab{}.
\newblock \showarticletitle{Investigating Evasive Techniques in SMS Spam Filtering: A Comparative Analysis of Machine Learning Models}.
\newblock \bibinfo{journal}{\emph{IEEE Access}} (\bibinfo{year}{2024}).
\newblock


\bibitem[Sjarif et~al\mbox{.}(2019)]%
        {sjarif2019sms}
\bibfield{author}{\bibinfo{person}{Nilam Nur~Amir Sjarif}, \bibinfo{person}{Nurulhuda Firdaus~Mohd Azmi}, \bibinfo{person}{Suriayati Chuprat}, \bibinfo{person}{Haslina~Md Sarkan}, \bibinfo{person}{Yazriwati Yahya}, {and} \bibinfo{person}{Suriani~Mohd Sam}.} \bibinfo{year}{2019}\natexlab{}.
\newblock \showarticletitle{SMS spam message detection using term frequency-inverse document frequency and random forest algorithm}.
\newblock \bibinfo{journal}{\emph{Procedia Computer Science}}  \bibinfo{volume}{161} (\bibinfo{year}{2019}), \bibinfo{pages}{509--515}.
\newblock


\bibitem[Tang et~al\mbox{.}(2022)]%
        {tang2022clues}
\bibfield{author}{\bibinfo{person}{Siyuan Tang}, \bibinfo{person}{Xianghang Mi}, \bibinfo{person}{Ying Li}, \bibinfo{person}{XiaoFeng Wang}, {and} \bibinfo{person}{Kai Chen}.} \bibinfo{year}{2022}\natexlab{}.
\newblock \showarticletitle{Clues in tweets: Twitter-guided discovery and analysis of SMS spam}. In \bibinfo{booktitle}{\emph{ACM CCS}}.
\newblock


\bibitem[Touvron et~al\mbox{.}(2023)]%
        {touvron2023llama}
\bibfield{author}{\bibinfo{person}{Hugo Touvron}, \bibinfo{person}{Thibaut Lavril}, \bibinfo{person}{Gautier Izacard}, \bibinfo{person}{Xavier Martinet}, \bibinfo{person}{Marie-Anne Lachaux}, \bibinfo{person}{Timoth{\'e}e Lacroix}, \bibinfo{person}{Baptiste Rozi{\`e}re}, \bibinfo{person}{Naman Goyal}, \bibinfo{person}{Eric Hambro}, \bibinfo{person}{Faisal Azhar}, {et~al\mbox{.}}} \bibinfo{year}{2023}\natexlab{}.
\newblock \showarticletitle{Llama: Open and efficient foundation language models}.
\newblock \bibinfo{journal}{\emph{arXiv preprint arXiv:2302.13971}} (\bibinfo{year}{2023}).
\newblock


\bibitem[Wei et~al\mbox{.}(2022)]%
        {wei2022chain}
\bibfield{author}{\bibinfo{person}{Jason Wei}, \bibinfo{person}{Xuezhi Wang}, \bibinfo{person}{Dale Schuurmans}, \bibinfo{person}{Maarten Bosma}, \bibinfo{person}{Fei Xia}, \bibinfo{person}{Ed Chi}, \bibinfo{person}{Quoc~V Le}, \bibinfo{person}{Denny Zhou}, {et~al\mbox{.}}} \bibinfo{year}{2022}\natexlab{}.
\newblock \showarticletitle{Chain-of-thought prompting elicits reasoning in large language models}.
\newblock \bibinfo{journal}{\emph{Advances in neural information processing systems}}  \bibinfo{volume}{35} (\bibinfo{year}{2022}), \bibinfo{pages}{24824--24837}.
\newblock


\bibitem[Yang et~al\mbox{.}(2024)]%
        {yang2024harnessing}
\bibfield{author}{\bibinfo{person}{Jingfeng Yang}, \bibinfo{person}{Hongye Jin}, \bibinfo{person}{Ruixiang Tang}, \bibinfo{person}{Xiaotian Han}, \bibinfo{person}{Qizhang Feng}, \bibinfo{person}{Haoming Jiang}, \bibinfo{person}{Shaochen Zhong}, \bibinfo{person}{Bing Yin}, {and} \bibinfo{person}{Xia Hu}.} \bibinfo{year}{2024}\natexlab{}.
\newblock \showarticletitle{Harnessing the power of llms in practice: A survey on chatgpt and beyond}.
\newblock \bibinfo{journal}{\emph{ACM Transactions on Knowledge Discovery from Data}} \bibinfo{volume}{18}, \bibinfo{number}{6} (\bibinfo{year}{2024}), \bibinfo{pages}{1--32}.
\newblock


\end{thebibliography}
\section*{Appendix}

\begin{table}[hbt]
\renewcommand{\arraystretch}{1.15}
\tabcolsep=0.05cm
\begin{center}
\caption{Performance evaluation of {\it LLMs} with Few-Shot Learning (examples based on varying lengths). Short and medium-length messages significantly enhance LLM performance in few-shot learning.} 
\label{tab:perform-message-length}
\begin{tabular}{ c | c | c | c | c | c | c }
 \hline
 & \multicolumn{6}{c}{\bf Performance Metrics}\\
\cline{2-7}
 \textbf{Model} & \textbf{Acc} & \textbf{FS} & \textbf{TPR} & \textbf{TNR} & \textbf{FPR} & \textbf{FNR} \\ [0.5ex]
 \hline\hline
 \multicolumn{7}{c}{Few-Shot Learning with short messages} \\
 \hline
LLAMA (70B) & 88.75\% & 88.57\% & 86.26\% & 91.30\% & 8.70\% & 13.74\% \\
LLAMA (13B) & 84.67\% & 86.40\% & 96.52\% & 72.60\% & 27.40\% & 3.48\% \\
\textbf{Mixtral (8x7B)} & \textbf{89.30\%} & \textbf{89.32\%} & \textbf{89.59\%} & \textbf{89.01\%} & \textbf{10.99\%} & \textbf{10.41\%} \\
Mixtral (7B) & 63.48\% & 63.15\% & 62.94\% & 38.72\% & 61.28\% & 27.06\% \\
DeepSeek (236B) & 93.50\% & 93.11\% & 87.89\% & 99.10\% & 0.90\% & 12.11\% \\
\textbf{GPT-4} & \textbf{96.80\%} & \textbf{96.70\%} & \textbf{93.99\%} & \textbf{99.60\%} & \textbf{0.40\%} & \textbf{6.01\%} \\
  \hline
 \multicolumn{7}{c}{Few-Shot Learning with Medium Size messages} \\
 \hline
\textbf{LLAMA (70B)} & \textbf{87.46\%} & \textbf{87.64\%} & \textbf{87.86\%} & \textbf{87.05\%} & \textbf{12.95\%} & \textbf{12.14\%} \\
LLAMA (13B) & 85.67\% & 87.08\% & 95.50\% & 75.61\% & 24.39\% & 4.50\% \\
Mixtral (8x7B) & 87.30\% & 87.66\% & 90.29\% & 84.32\% & 15.68\% & 9.71\% \\
Mixtral (7B) & 74.29\% & 74.15\% & 73.94\% & 37.28\% & 62.72\% & 26.06\% \\
DeepSeek (236B) & 95.10\% & 94.93\% & 91.89\% & 98.30\% & 1.70\% & 8.11\% \\
\textbf{GPT-4} & \textbf{97.18\%} & \textbf{97.03\%} & \textbf{95.03\%} & \textbf{99.20\%} & \textbf{0.80\%} & \textbf{4.97\%} \\
  \hline
   \multicolumn{7}{c}{Few-Shot Learning with Lengthy messages} \\
 \hline
\textbf{LLAMA (70B)} & \textbf{74.95\%} & \textbf{75.09\%} & \textbf{75.58\%} & \textbf{74.33\%} & \textbf{25.67\%} & \textbf{24.42\%} \\
LLAMA (13B) & 72.83\% & 71.29\% & 76.26\% & 70.11\% & 29.89\% & 23.74\% \\
Mixtral (8x7B) & 69.40\% & 61.70\% & 49.35\% & 89.41\% & 10.59\% & 50.65\% \\
Mixtral (7B) & 63.45\% & 63.12\% & 62.90\% & 36.78\% & 63.22\% & 30.10\% \\
DeepSeek (236B) & 81.15\% & 76.94\% & 62.96\% & 99.30\% & 0.70\% & 37.04\% \\
\textbf{GPT-4} & \textbf{94.10\%} & \textbf{93.73\%} & \textbf{88.29\%} & \textbf{99.90\%} & \textbf{0.10\%} & \textbf{11.71\%} \\
  \hline
\end{tabular}
\end{center}
\end{table}

\begin{table*}[ht]
\caption{Attack methods and their descriptions.}
\centering
\label{tab:attacks}
\begin{tabular}{>{\centering\arraybackslash}p{1.1cm}|p{1.65cm}|p{14cm}}
\toprule
\hline
\textbf{Serial \#} & \textbf{Attack} & \textbf{Description} \\ \hline
\multicolumn{3}{c}{\textbf{Perceptible or Conventional Attacks}} \\ \hline
a. & Spacing & Insert a space in the word. \\ \hline
b. & Delete Chars & Delete a random character of the word except for the first. \\ \hline
c. & Swap Chars & Swap two adjacent letters in the word at random, ensuring the first and last letters remain unchanged. \\ \hline
d. & Insert Chars & Insert a random letter into the word. \\ \hline
e. & Sub-Chars & Substitute characters with visually similar alternatives (e.g., replacing ``l'' with ``1'', ``s'' with ``\$'', ``o'' with ``0''). \\ \hline
f. & Sub-Word & Substitute a word with one of its top-k closest neighbors based on a context-aware word vector space model. \\ \hline
\multicolumn{3}{c}{\textbf{Imperceptible Attacks}} \\ \hline
g. & Invisible & Insert characters that are invisible to human eyes (e.g., zero-width spaces), altering text processing without changing visual appearance. \\ \hline
h. & Homoglyphs & Use characters from different scripts that look similar to commonly used characters (e.g., replacing Latin ``a'' with Cyrillic ``a'') to confuse text processing systems and mislead users. \\ \hline
i. & Reorderings & Manipulate the visual rendering of text using bidirectional control characters such as Right-to-Left Override (RTLO), causing the text to be displayed in reverse order. \\ \hline
\bottomrule
\end{tabular}
\end{table*}

\begin{table*}[hbt!]
\renewcommand{\arraystretch}{1.15}
\tabcolsep=0.05cm
\begin{center}
\caption{Robustness evaluation of {\it LLMs (Pre-trained vs Fine-tuned)}. Most pre-trained and fine-tuned LLMs exhibit strong robustness against adversarial attacks. Variability across models underscores the importance of both model selection and fine-tuning.} 
\label{tab:robust_llms}
\scalebox{0.9}{
\begin{tabular}{c|p{1.3cm}|p{1.3cm}|p{1.1cm}|p{1.1cm}|p{1.1cm}|p{1.1cm}|p{1.1cm}|p{1.5cm}|p{1.1cm}|p{1.3cm}}
\toprule
 \hline
 & \multicolumn{9}{c}{\bf Adversarial Attacks}\\
\cline{3-11}
 \textbf{Model} & \textbf{Original}&\textbf{Spacing} & \textbf{Insert Chars} &\textbf{Delete Chars} &\textbf{Sub-Chars} &\textbf{Sub Word} &\textbf{Swap Chars} &\textbf{Invisible} &\textbf{Homo} &\textbf{Reorder} \\ [0.5ex]
 \hline
   \multicolumn{7}{c}{Pre-trained LLMs} \\
 \hline
LLAMA (70B) & 96.9\% & 96.9\% & 98.5\% & 96.9\% & 98.5\% & 97\% & 96.9\% & 98.5\% & 94.9\% & 97.2\% \\
LLAMA (13B) & 94.9\% & 96.4\%  & 97.9\% & 96.9\% & 98.2\% & 97.4\% & 97.9\% & 96.4\% & 96.9\% & 96.4\% \\
Mixtral (70B) & 95.9\% & 98.9\% & 99\% & 98.9\% & 99.5\% & 97.2\% & 99.5\% & 98.2\% & 96.9\% & 98.2\% \\
Mistral (7B) & 94.9\% & 96.9\% & 96.9\% & 96.9\% & 96.9\% & 94.9\% & 96.9\% & 97\% & 94.9\% & 97\% \\
DeepSeek & 89.3\%  & 91.5\% & 94.9\% & 93.2\% & 94.9\% & 92.9\% & 94.5\% & 81.6\% & 90.5\% & 86.2\% \\
GPT-4 & 98.7\% & 98.7\% & 99.7\% & 99.7\% & 99.7\% & 99.5\% & 99.7\% & 99.2\% & 99.2\% & 98.9\% \\
 \hline
   \multicolumn{7}{c}{Fine-tuned LLMs} \\
 \hline
LLAMA (70B) & 97.2\% & 97.9\% & 98.5\% & 98.9\% & 98\% & 98.5\% & 96.9\% & 98.5\% & 97.9\% & 98.9\% \\
LLAMA (13B) & 96.4\% & 97\% & 97.2\% & 97.4\% & 97.4\% & 98\% & 97.2\% & 97.9\% & 97.4\% & 98.2\% \\
Mixtral (8x7B) & 98.7\% & 100\%  & 100\% & 100\% & 98.9\% & 99.2\% & 100\% & 97.7\% & 97.4\% & 98.5\% \\
  \bottomrule
\end{tabular}}
\end{center}
\end{table*}

\begin{figure*}[th]
\centering
\includegraphics[width=0.8\linewidth]{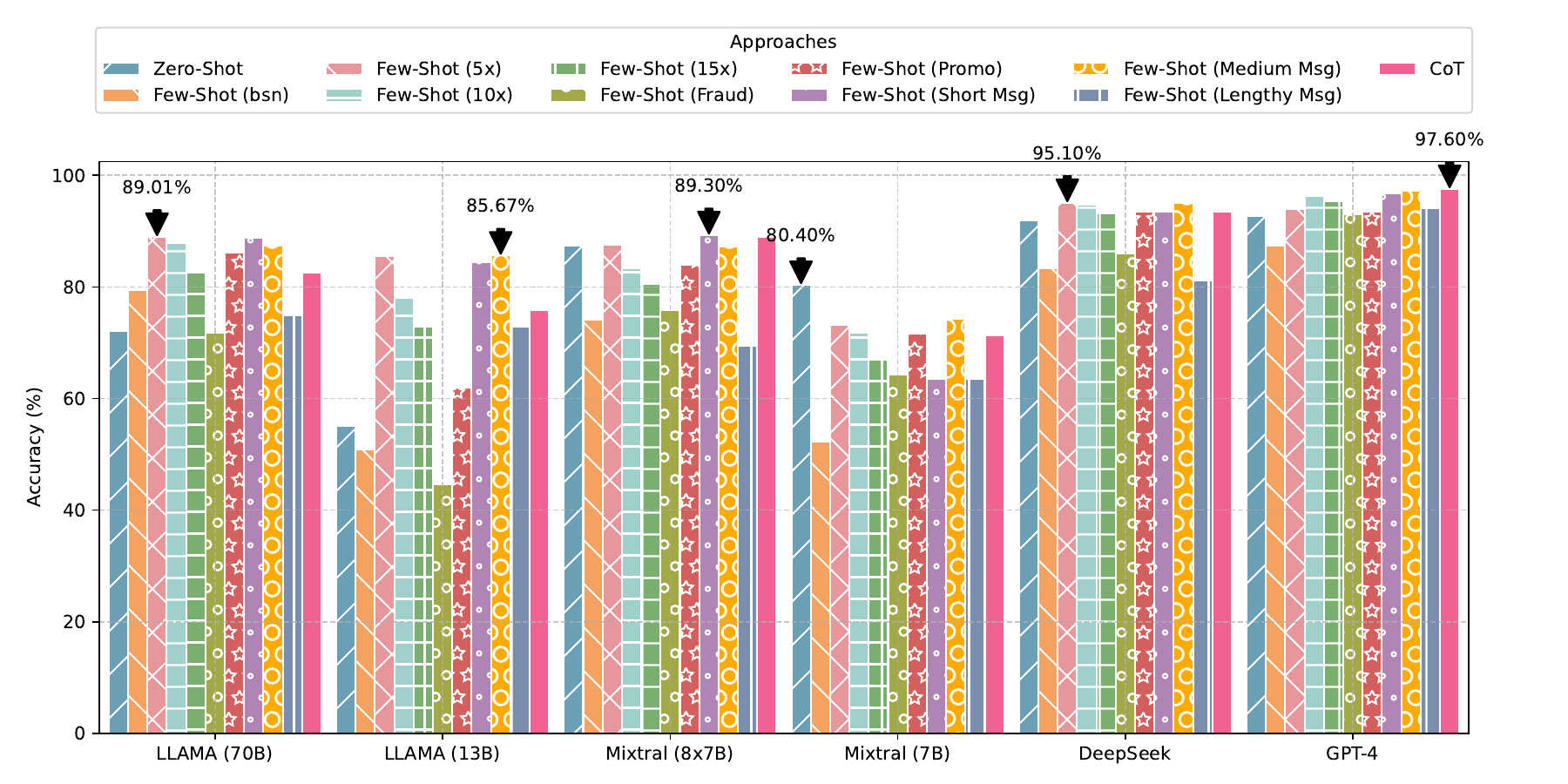}
\caption{Accuracy of LLMs across various learning approaches. The highest accuracy for each LLM is indicated by an arrow and labeled with the achieved accuracy. Abbreviations: bsn (baseline), Msg (messages), CoT (chain-of-thought)}
\label{fig:ablation}
\end{figure*}

\begin{table}[hbt!]
\renewcommand{\arraystretch}{1.15}
\tabcolsep=0.05cm
\begin{center}
\caption{Performance evaluation of {\it LLMs} with Chain-of-thought. Chain-of-thought prompting improves reasoning abilities, with varying effectiveness across models.}
\label{tab:perform-chain}
\begin{tabular}{ c | c | c | c | c | c | c }
\toprule
 & \multicolumn{6}{c}{\bf Performance Metrics}\\
\cline{2-7}
 \textbf{Model} & \textbf{Acc} & \textbf{FS} & \textbf{TPR} & \textbf{TNR} & \textbf{FPR} & \textbf{FNR} \\ [0.5ex]
 \midrule
  \hline
  LLAMA (70B) & 82.51\% & 84.22\% & 92.48\% & 72.35\% & 27.65\% & 7.52\% \\
  LLAMA (13B) & 78.69\% & 77.83\% & 97.39\% & 59.21\% & 40.79\% & 2.61\% \\
 \textbf{Mixtral (8x7B)} & \textbf{88.89\%} & \textbf{89.76\%} & \textbf{94.75\%} & \textbf{82.71\%} & \textbf{17.29\%} & \textbf{5.25\%} \\
  Mixtral (7B) & 71.27\% & 73.15\% & 77.08\% & 65.27\% & 34.73\% & 22.92\% \\
  DeepSeek (236B) & 93.45\% & 93.24\% & 90.39\% & 96.50\% & 3.50\% & 9.61\% \\
 \textbf{GPT-4 (1.76T)} & \textbf{97.60\%} & \textbf{97.60\%} & \textbf{96.10\%} & \textbf{99.10\%} & \textbf{0.90\%} & \textbf{3.90\%} \\
  \bottomrule
\end{tabular}
\end{center}
\end{table}

\subsection*{Chain-of-Thought Prompting.}
\label{sec:cot_appendix}

The chain of thought prompt is structured to ensure a precise and unbiased evaluation of each SMS, isolating each analysis to prevent influence from previous evaluations. The process begins with a detailed linguistic review of the SMS, correcting any typographical and grammatical errors and identifying slang, which establishes a clear baseline for understanding the message's intent.

The analysis then progresses to contextual examination, where specific and verifiable details indicative of legitimate offers are sought. This step is crucial for distinguishing authentic messages from potential scams. Concurrently, the main topics of the SMS are identified to encapsulate the core message content, and any social engineering tactics are noted.

Key components such as URLs, callback phone numbers, and follow-up actions demanding financial transactions or personal details are extracted and scrutinized. The language used is analyzed for promotional content and typical scam indicators, including exaggerated urgency, too-good-to-be-true offers, and manipulative language aimed at eliciting immediate action.

The sender's intent is critically assessed, with checks for authenticity through recognizable brand mentions and the level of message personalization. URLs are particularly examined for their legitimacy, especially those linked to financial transactions or sensitive information.

This systematic evaluation culminates in a definitive classification of the SMS as benign (0) or spam (1), ensuring a comprehensive and logical flow from initial linguistic correction to final decision-making based on the content's authenticity and intent.

\subsection*{Performance Categorization}
\label{sec:spam_catg_appendix}

\begin{itemize}[leftmargin=*]
    \item \textbf{Satisfactory:} In the \textit{Satisfactory} category, the system achieves an FPR and FNR both below 5\%. This indicates a functional system where the majority of spam is correctly identified, and most legitimate messages are delivered as expected. However, users may still experience occasional false positives, where legitimate messages are mistakenly flagged as spam, or false negatives, where spam messages are missed and delivered to the user's inbox. While this level is acceptable for general use, there may still be room for improvement, especially in environments where the cost of missing legitimate messages is high.

    \item \textbf{Good:} The \textit{Good} category is characterized by an FPR and FNR both below 3\%. Systems within this range demonstrate a higher level of accuracy, with fewer instances of incorrect classification. Users experience minimal disruption due to false positives, and the system is effective at filtering out the majority of spam messages. This level of performance is generally well-received in practice, offering a solid balance between security and user convenience.

    \item \textbf{Perfect:} At the \textit{Perfect} level, both FPR and FNR are below 1\%. This represents the ideal performance scenario, where the system almost never misclassifies messages. Users can trust that legitimate messages will not be flagged as spam, and that spam messages will be reliably filtered out. Achieving this level is particularly challenging, especially in scenarios involving few-shot learning, but it is the benchmark for optimal user satisfaction and security.
\end{itemize}

\subsection*{Prompt Design}
\label{lab:prompt}

We designed simple prompts following the best practices provided by OpenAI \cite{prompteng} to instruct the LLM in determining the label of an SMS for spam classification. The prompts for zero-shot, few-shot, and Chain-of-Thought are shown in Listings \ref{fig:zero_shot_prompt}, \ref{fig:few_shot_prompt}, and \ref{fig:cot_prompt}, respectively.

\begin{lstlisting}[caption={Zero-shot Prompt: A simple instruction to classify SMS as 'Spam' or 'Ham' without examples.}, label={fig:zero_shot_prompt}]
{
    "role": "system",
    "content": "Act as an SMS spam classifier. Your task is to classify whether a given SMS is 'Spam' or 'Ham' (benign). Do not reprint the SMS or provide any explanation; simply respond with 'Spam' for spam messages or 'Ham' for benign messages."
}
\end{lstlisting}

\begin{lstlisting}[caption={Few-shot Prompt: The prompt includes examples of 'Spam' and 'Ham' messages to guide the model.}, label={fig:few_shot_prompt}]
{
    "role": "system",
    "content": 
"""Act as an SMS spam classifier. Your task is to classify whether a given SMS is 'Spam' or 'Ham' (benign). Do not reprint the SMS or provide any explanation; simply respond with 'Spam' for spam messages or 'Ham' for benign messages.
    Below are some examples of 'spam' and 'ham' Messages:
    For example: the output for 'Message1' is Spam.
    the output for 'Message2' is Spam.
    .......
    .......
    the output for 'Message N' is Ham. """
}
\end{lstlisting}
%
%

\begin{lstlisting}[caption={Chain-of-Thought Prompt: A detailed prompt guiding the model through a comprehensive analysis before classification.}, label={fig:cot_prompt}]
{
  "role": "system",
  "content": """You are an SMS spam Analyst API capable of dialogue analysis.
  Your task is to classify whether a given SMS is spam or benign. Do not reprint the SMS nor give any explanation; just answer with 'Spam' for a spam message or 'Ham' for a benign message.
  Begin each new analysis type with a fresh perspective, avoiding influence from scores assigned in the previous analysis. Ensure that your evaluation of SMS text does not influence or is influenced by the scores of other SMS texts.
  Thoroughly assess the text of the SMS. Process the language, identify slang expressions, and recover typing and grammar mistakes.  
  Identify the context of the SMS, looking for specific and verifiable details typical of legitimate offers. Identify the main topics in the SMS.
  Identify any social engineering tactics involved, if any. Extract any URL and callback phone number.
  Identify the follow-up actions present, if any, scrutinizing for immediate financial transactions or provided contact details.
  Recognize promotional and advertisement language. Detect potential scam indicators, such as urgency, too-good-to-be-true offers, suspicious links, and requests for personal information.
  Carefully analyze the sender's intent and context. Detect SMS that might represent legitimate offers, promotions, or advertisements and differentiate them from scams.
  Consider how the message might exploit human psychology or prompt immediate action.
  Check for exaggerated offers and urgency: Identify phrases that create a sense of urgency and offer unbelievable deals.  
  Look for persuasive and emotionally charged language: Detect emotionally manipulative language aimed at prompting immediate action.  
  Verify source authenticity: Check if the message mentions a recognizable and trusted brand or source.  
  Assess personalization: Determine if the message uses personal details or is generic and impersonal.  
  Evaluate the presence of URLs and links: Identify and scrutinize URLs for authenticity.  
  Consider the nature of financial transactions and sensitive information: Be cautious with messages involving financial details or sensitive information requests.  
  Based on the above analysis, determine whether the message is 'ham' or 'spam'.
  Below are some examples of 'spam' and 'ham' Messages:
  For example: the output for 'Message1' is spam.
  The output for 'Message2' is spam.
  .......
  .......
  The output for 'Message N' is ham."""
}
\end{lstlisting}

\end{document}